\def\year{2021}\relax
\documentclass[letterpaper]{article} % DO NOT CHANGE THIS
\usepackage{aaai21}  % DO NOT CHANGE THIS
\usepackage{times}  % DO NOT CHANGE THIS
\usepackage{helvet} % DO NOT CHANGE THIS
\usepackage{courier}  % DO NOT CHANGE THIS
\usepackage[hyphens]{url}  % DO NOT CHANGE THIS
\usepackage{graphicx} % DO NOT CHANGE THIS
\usepackage{natbib}  % DO NOT CHANGE THIS AND DO NOT ADD ANY OPTIONS TO IT
\usepackage{caption} % DO NOT CHANGE THIS AND DO NOT ADD ANY OPTIONS TO IT
\newcommand{\figdir}{figs}
\def\swone{0.98\linewidth}

\def\swfive{0.17\linewidth}
\def\swsix{0.14\linewidth}

\def\swnine{0.09\linewidth}

\newcommand{\reffig}[1]{Figure.~\ref{#1}}

\newcommand{\reftable}[1]{Table.~\ref{#1}}

\title{Efficient Deep Image Denoising via Class Specific Convolution}
\author {
	Lu Xu\textsuperscript{\rm 1}\thanks{Equal contribution}\thanks{This work was done when Lu Xu was an intern at SenseTime.},
	Jiawei Zhang\textsuperscript{\rm 2}\footnotemark[1]\thanks{Corresponding author},
	Xuanye Cheng\textsuperscript{\rm 2}, 
	Feng Zhang\textsuperscript{\rm 2},
	Xing Wei\textsuperscript{\rm 3},
	Jimmy Ren\textsuperscript{\rm 2, \rm 4}
	\\
}

\affiliations {
	\textsuperscript{\rm 1} The Chinese University of Hong Kong \\
	\textsuperscript{\rm 2} SenseTime Research \\
	\textsuperscript{\rm 3} School of Software Engineering, Xi'an Jiaotong University \\
	\textsuperscript{\rm 4} Qing Yuan Research Institute, Shanghai Jiao Tong University \\
	lxu7@cse.cuhk.edu.hk, zhjw1988@gmail.com, xuanyech@gmail.com,\\ gleefeng@foxmail.com, weixing@mail.xjtu.edu.cn, jimmy.sj.ren@gmail.com
}
\begin{document}

	\maketitle
%	\linenumbers
	
	\begin{abstract}
		Deep neural networks have been widely used in image denoising during the past few years.
		Even though they achieve great success on this problem, they are computationally inefficient which makes them inappropriate to be implemented in mobile devices.
		In this paper, we propose an efficient deep neural network for image denoising based on pixel-wise classification \footnote{\url{https://github.com/XenonLamb/CSConvNet}}.
		Despite using a computationally efficient network cannot effectively remove the noises from any content, it is still capable to denoise from a specific type of pattern or texture.
		The proposed method follows such a divide and conquer scheme.
		We first use an efficient U-net to pixel-wisely classify pixels in the noisy image based on the local gradient statistics.
		Then we replace part of the convolution layers in existing denoising networks by the proposed Class Specific Convolution layers (CSConv) which use different weights for different classes of pixels.
		Quantitative and qualitative evaluations on public datasets demonstrate that the proposed method can reduce the computational costs without sacrificing the performance compared to state-of-the-art algorithms.

		%\keywords{Denoising, Classification, Efficiency}
	\end{abstract}

	\section{Introduction}
	
	Image denoising aims to recover a clean image $X$ given a noisy observation $Y$ which is one of the most fundamental problems in low-level vision.
	% 
	%	Assuming the noise $N$ is additive, the noisy observation $Y$ can be modeled as:
	%	%
	%	\begin{equation}
	%	\label{eq:noisemodel}
	%	Y=X+N.
	%	%\vspace{-1mm}
	%	\end{equation}
	%
	%in which $X$ is the clean image which should be restored.
	%
	Plenty of algorithms~\cite{buades2005non, elad2006image, dabov2007image, zoran2011learning, dong2012nonlocally, gu2014weighted} have been proposed to solve this problem in the last decade.
	Even though they are effective to remove noises, their computational costs are high because of the complex optimization process as well as block matching procedure which make them inappropriate to be deployed in mobile devices.

	Recently, deep neural networks have been widely used in image denoising \cite{mao2016image, zhang2017beyond, tai2017memnet, jia2019focnet, gu2019self, liu2018non, zhang2019residual, brooks2019unprocessing, zamir2020cycleisp}.
	To achieve state-of-the-art performance, very deep network structures with residual net~\cite{zhang2017beyond}, dense net~\cite{tai2017memnet, jia2019focnet} and U-net~\cite{mao2016image} are applied.
	As a result, they are still computationally expensive.
	In order to make the network more efficient,  \cite{gu2019self} use a multi-scale structure and the low-resolution intermediate features can save the time cost as well as the memory.
	In super-resolution, \cite{ahn2018fast} replace the convolution by group convolution followed with pointwise one to accelerate the inference time similar to MobileNet~\cite{howard2017mobilenets}.

	Nowadays, knowledge distilling~\cite{hinton2014distilling}, parameter pruning~\cite{han2015learning} and network quantization~\cite{jacob2018quantization} are widely used to compress the network.
	However, these techniques are commonly used in high-level vision tasks and only a few works, \emph{e.g.} \cite{hui2018fast}, have discussed applying them to low-level vision.
	%
	%	The reason may come from that the deep neural network based low-level vision methods are a regression task and require pixel-wise accuracy estimation, any techniques that reduce the model size or weights accuracy may have a direct influence on the restoration performance.
	%
	
	\begin{figure*}[!t]\footnotesize
		\begin{center}
			\begin{tabular}{c}
				\includegraphics[width=\swone]{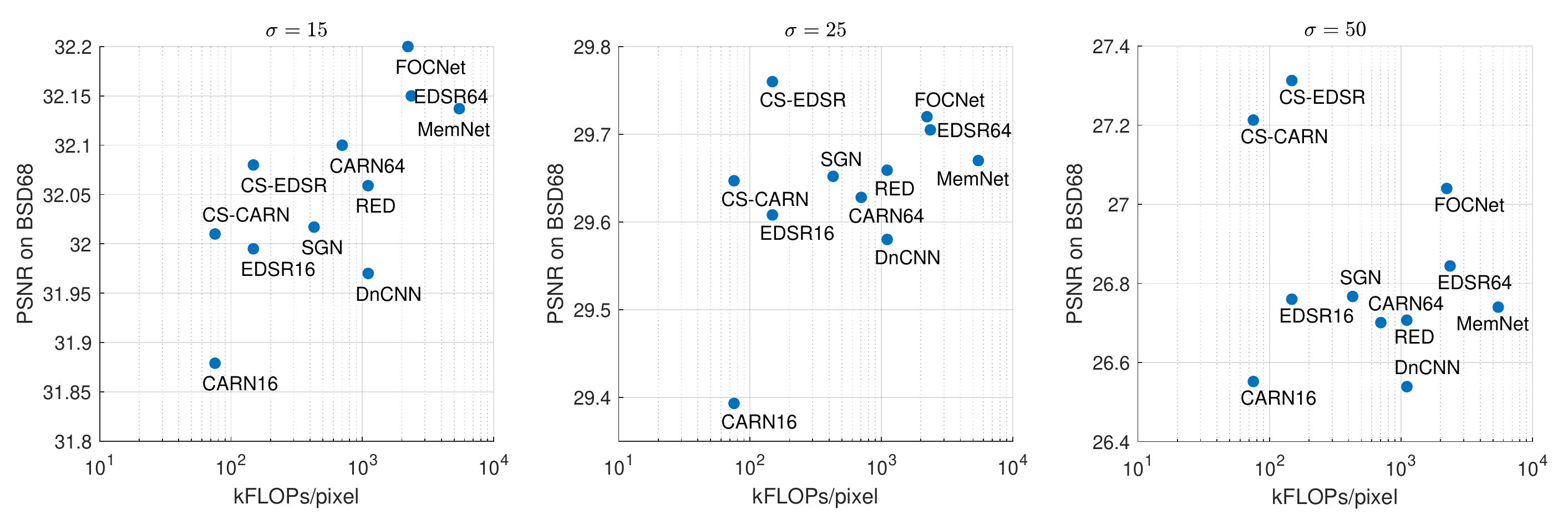} \\
			\end{tabular}
		\end{center}
		%\vspace{-3mm}
		\caption{
			Comparison of PSNR and FLOPs per pixel: the proposed networks CS-EDSR and CS-CARN with existing image denoising methods such as EDSR, CARN, RED, FOCNet, DnCNN, MemNet, SGN under $\sigma = $ 15, 25 and 50 noises.
			EDSR64 and CARN64 are the original EDSR and CARN structure with 64 channels.
			EDSR16, CARN16, CS-EDSR and CS-CARN only contain 16 channels.
			It shows that the proposed CS-EDSR and CS-CARN perform comparable relative to other methods while requiring much fewer FLOPs per pixel.
			In addition, the PSNR improves by a large margin between EDSR16 and CS-EDSR as well as CARN16 and CS-CARN which demonstrates the effectiveness of the proposed CSConv.
		}
		\label{fig:flops}
	\end{figure*}
	
	Even though using a small network to denoise in any scenarios is difficult, it is still possible to remove noises from only a specific type of pattern or texture.
	Such a divide and conquer scheme has already been applied in low-level vision tasks.
	RAISR~\cite{romano2016raisr} and BLADE~\cite{getreuer2018blade} classify the image patches into different buckets according to the local gradient statistics.
	Then only one specific convolution layer is learned for every bucket to efficiently solve low-level vision tasks.
	Kernel prediction network (KPN) is proposed to remove noises from brust images \cite{mildenhall2018burst} or single-frame image \cite{bako2017kernel, vogels2018denoising}.
	Unlike RAISR and BLADE that classify patches, they use deep neural networks to predict spatially variant kernels which not only align the frames but also remove noises according to the specific information around every noisy pixel.
	\cite{wang2018recovering} propose spatial feature transform (SFT) for super-resolution.
	%
	%Different from \cite{mildenhall2018burst} applied spatially variant convolution in images, SFT, which is similar to an $1\times1$ KPN, is applied in features and the transforms are predicted from semantic information.
	They use the semantic information to decide the spatial feature transforms to generate realistic textures.
	
	However, shortcomings still exist in RAISR, BLADE and KPN.
	RAISR and BLADE are equivalent to a single layer network which is too shallow to remove severe noises.
	In addition, the local gradient characteristics estimated from eigenanalysis are not very accurate under noises.
	Although KPN estimates spatially variant kernels by a network that is trained from the training set, it confronts the same issue that the spatially variant convolution is directly applied to images which is too shallow to remove severe noises from a single image.
	Also, KPN is a large U-net~\cite{ronneberger2015u} to estimate the pixel-wise spatially variant kernels.
	%
	%	It is difficult to extend KPN in feature domain of a deep neural network as the number of features, which is $C_{in}\times C_{out} \times K^2$ where $C_{in}$, $C_{out}$ and $K$ are the number of input features, the number of output features and the kernel size of the spatially variant convolution layer, estimated from KPN are too large.
	It is even difficult to extend KPN in feature domain of a deep neural network as the number of features\footnote{The number of features is $C_{in}\times C_{out} \times K^2$ where $C_{in}$, $C_{out}$ and $K$ are the number of input features, the number of output features and the kernel size of the spatially variant convolution.} estimated from KPN are too large.

	In this paper, we propose a class specific convolution (CSConv) to replace part of the convolutions in existing image restoration networks, \emph{e.g.} EDSR~\cite{lim2017enhanced} and CARN~\cite{ahn2018fast}, to efficiently remove noises from images in the proposed CSConv-based denoising convolutional network (CSDN).
	As can be seen in \reffig{fig:flops}, we compare the PSNR and FLOPs per pixel by the proposed network with state-of-the-art ones.
	It shows that the network with the proposed CSConv can reduce the FLOPs while maintaining the denoising performance.
	We follow the divide and conquer scheme like RAISR and BLADE to classify the noisy pixels into different classes.
	Even though their classification is effective, it fails especially when the input image is too noisy as in \reffig{fig:effective}(g).
	To more accurately classify the pixels, we propose a pixel-wise classification network (PCN) which uses an efficient U-net other than eigenanalysis in RAISR and BLADE.
	Although the inputs are the noisy pixels, the labels of PCN are estimated from the clean images based on the eigenanalysis in RAISR and BLADE which are shown in \reffig{fig:effective}(f).
	As PCN does not need to estimate the pixel-wise kernels as in KPN, it can apply a more computationally efficient U-net in the proposed PCN.
	Specifically, we adopt group convolutions to reduce the computational costs and the time cost for classification can almost be ignored relative to the following CSDN.
	As can be seen in \reffig{fig:effective}(h), the proposed PCN can predict more accurate gradient statistics which will lead to a better pixel-wise classification.
	Then, we can use a more efficient network to remove the noises from every class in CSDN.
	Specifically, we directly reduce the number of features in EDSR and CARN by four times which means only about $1/16$ FLOPs are needed relative to the original baseline networks.
	Both the quantitative and qualitative experiments demonstrate that the computational efficient networks with the proposed CSConv can perform favorably against the original baseline networks.
	The proposed CSConv is flexible and can be adopted into almost any existing state-of-the-art denoising networks to reduce their computational costs while maintaining the performance.
	%
	
	%	%
	%	\begin{figure*}[!t]\footnotesize
	%		\begin{center}
	%			\begin{tabular}{c}
	%				\includegraphics[width=\swone]{\figdir/csdn.pdf} \\
	%			\end{tabular}
	%		\end{center}
	%		%\vspace{-3mm}
	%		\caption{
	%			CS-EDSR architecture for class-specific denoising by replacing some convolution layers with CSConv in EDSR.
	%			%
	%			RB denotes Residual Block. CSRB denotes Class Specific Residual Block. 
	%			%
	%			$\bigoplus$ denotes element-wise addition, and $\bigotimes$ denotes convolution operation.
	%			%
	%			In the CSConv module, $W_1$, $W_2$, ..., $W_M$ denote the filters of size $C_{in} \times C_{out} \times K^2$ for class 1,2, ..., M. 
	%			%
	%			Pixel-wise classes select one filter for each position to apply convolution on the $K \times K$ region around that position according to \refeq{eq:csconv}. }
	%		\label{fig:csdn}
	%	\end{figure*}
	
	%
	
	The contributions of the proposed methods can be summarized as follows:
	\begin{itemize}
		\item The proposed CSDN considers image denoising in a divide and conquer scheme by the proposed CSConv where different weights will be applied to different classes of pixels and an efficient network is used to remove noises.
		\item Unlike estimating spatially variant kernels by a computational inefficient network, the pixel classification is easy to learn with an efficient network.
		\item The experimental results demonstrate that the proposed network can perform favorably against state-of-the-art methods with much less computational costs. 
	\end{itemize}

	\begin{figure*}[!t]\footnotesize
		\begin{center}
			\begin{tabular}{c}
				\includegraphics[width=\swone]{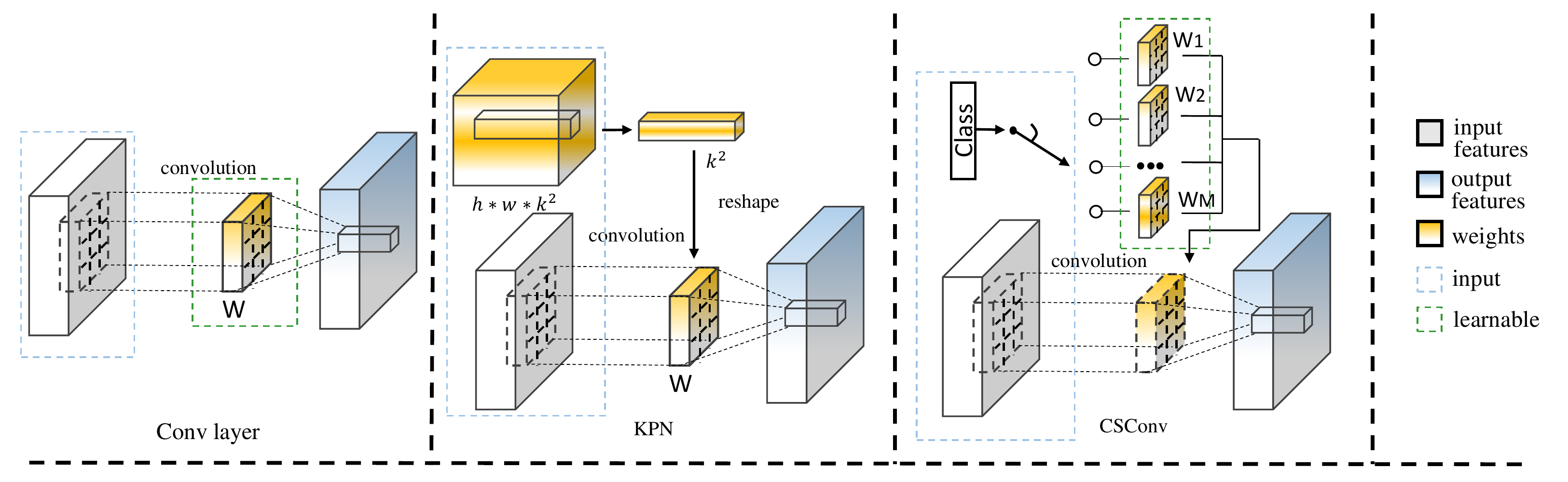} \\
			\end{tabular}
		\end{center}
		%\vspace{-30mm}
		\caption{
			Comparison between the proposed CSConv with convolution layer and KPN.
			In convolution layer, all the spatial positions share the same weights $W$.
			KPN and CSConv apply different $W$ in different locations.
			However, $W$ is predicted from a large network in KPN.
			In the proposed CSConv, it uses PCN to classify pixels into different classes and applies different $W$ according to the classification which is more efficient.
			Similar to convolution layer, ${W_1, W_2, ..., W_M}$ are learned in the training process in CSConv.
		}
		\label{fig:csconv}
	\end{figure*}	
	
	\section{Method}
	In this section, we will describe the details of the proposed pipeline which consists of two subnetworks in sequence: a lightweighted pixel-wise classification network (PCN) and a CSConv-based denoising network (CSDN).

	The key idea of this pipeline is utilizing gradient statistics to apply the divide-and-conquer strategy in which only a small network is used to remove noises from a specific class of pattern or texture. 
	Given a noisy image $Y$, PCN produces a pixel-wise map of noise-free image gradient statistics $(\hat{\phi},\hat{\lambda},\hat{\mu}) = \mathbf{F}_{C}(Y;\theta_{C})$, where $\theta_{C}$ denotes the network parameters of PCN, $\hat{\phi}$, $\hat{\lambda}$ and $\hat{\mu}$ are the estimated gradient orientation, strength and coherence.
	The gradient statistics map is further quantized into a class map $\hat{H} = \mathbf{H}(\hat{\phi},\hat{\lambda},\hat{\mu})$ using the hash table in RAISR.
	%%%%%%%%%%%% should we add the details about the hash process????
	%
	Elements of $\hat{H}$ are pixel-wise class indices in range $\{1, 2,..., M\}, M= M_{\phi} \times M_{\lambda} \times M_{\mu}$ which denotes the total number of classes used in CSConv and $M_{\phi}, M_{\lambda}, M_{\mu}$ are the numbers of classes for $\phi, \lambda$ and $\mu$ in the hash table.
	Then, CSDN restores the denoised image $\hat{X} = \mathbf{F}_{D}(Y,H;\theta_{D})$, where $\theta_{D}$ denotes the network parameters of CSDN.
	The details are described in the following subsections.
	
	%	We will demonstrate how to utilize PCN to classify the noisy pixels in \refsec{sec: pcn}.
	%	%
	%	Then, the proposed class specific convolution (CSConv) will be described in \refsec{sec: csconv} and how to use it in CSDN will be shown in \refsec{sec: csdn}.
	%	%
	%	At last, the loss functions to train PCN and CSDN will be described in \refsec{sec: loss}.
	
	%    We will demonstrate the proposed PCN and CSDN in details in the following subsections.
	
	\begin{figure*}[!t]
		\begin{center}
			\begin{tabular}{ccccc}
				\includegraphics[width=\swfive]{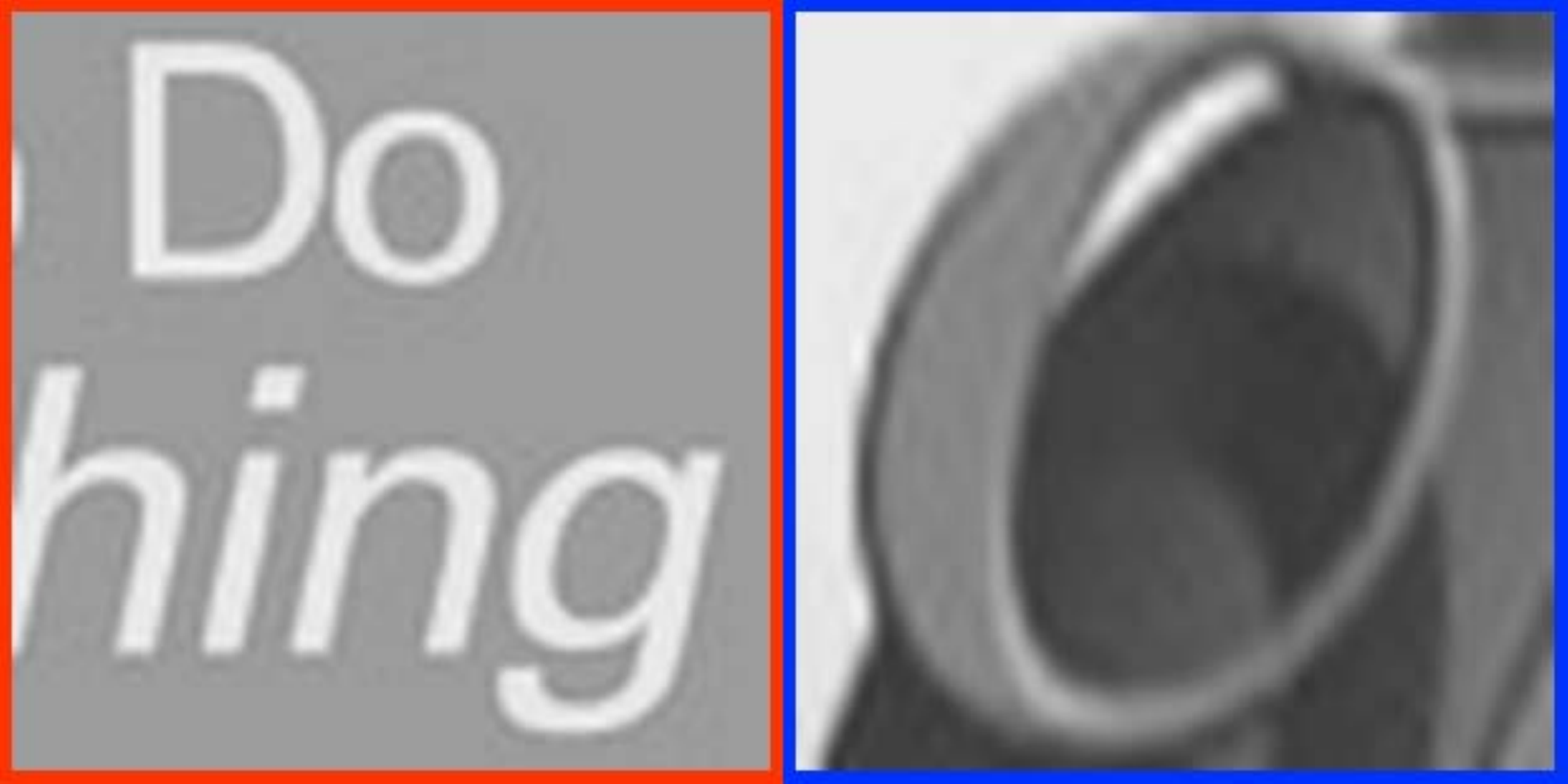} &
				\includegraphics[width=\swfive]{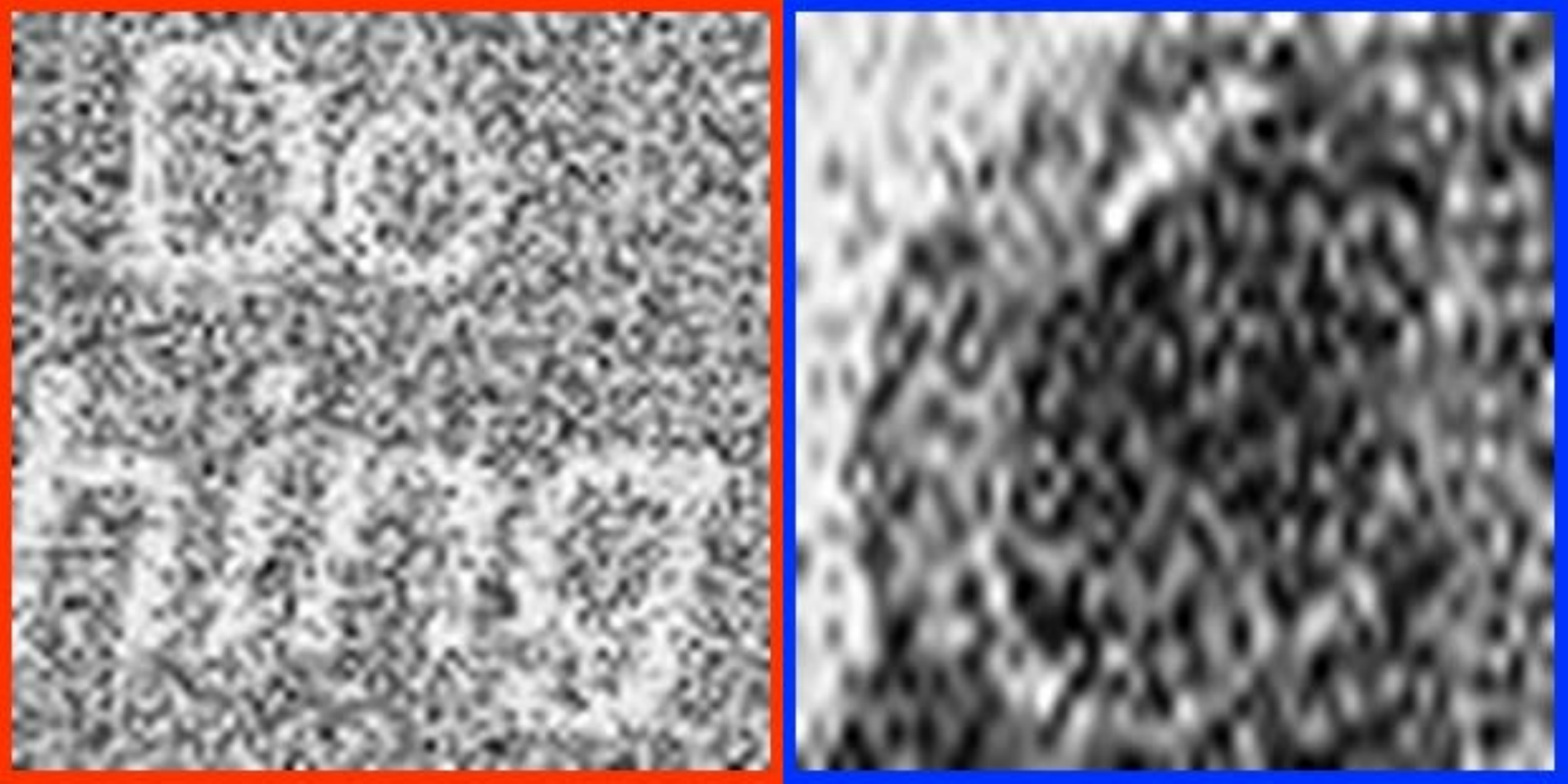} &
				\includegraphics[width=\swfive]{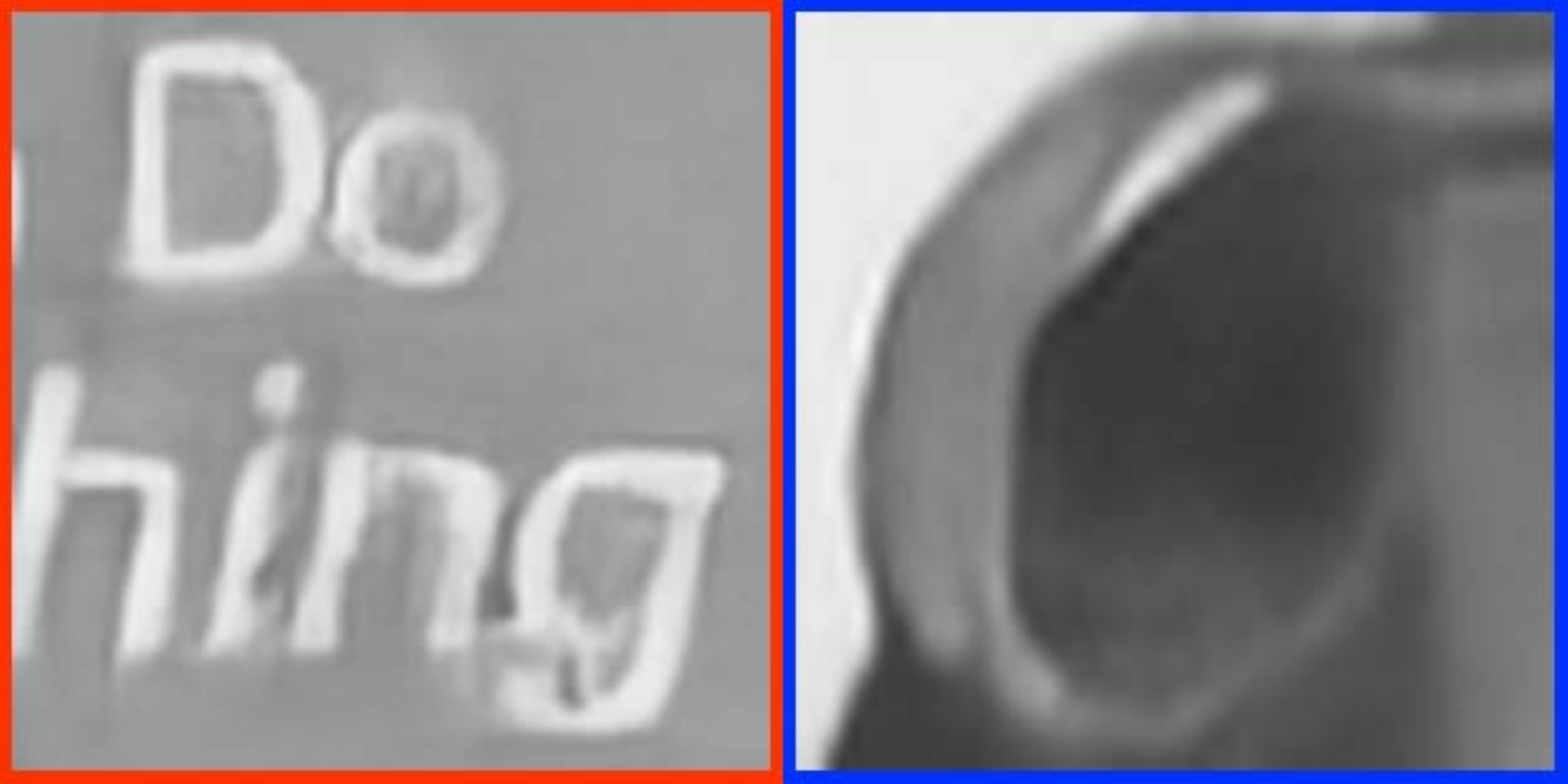} &
				\includegraphics[width=\swfive]{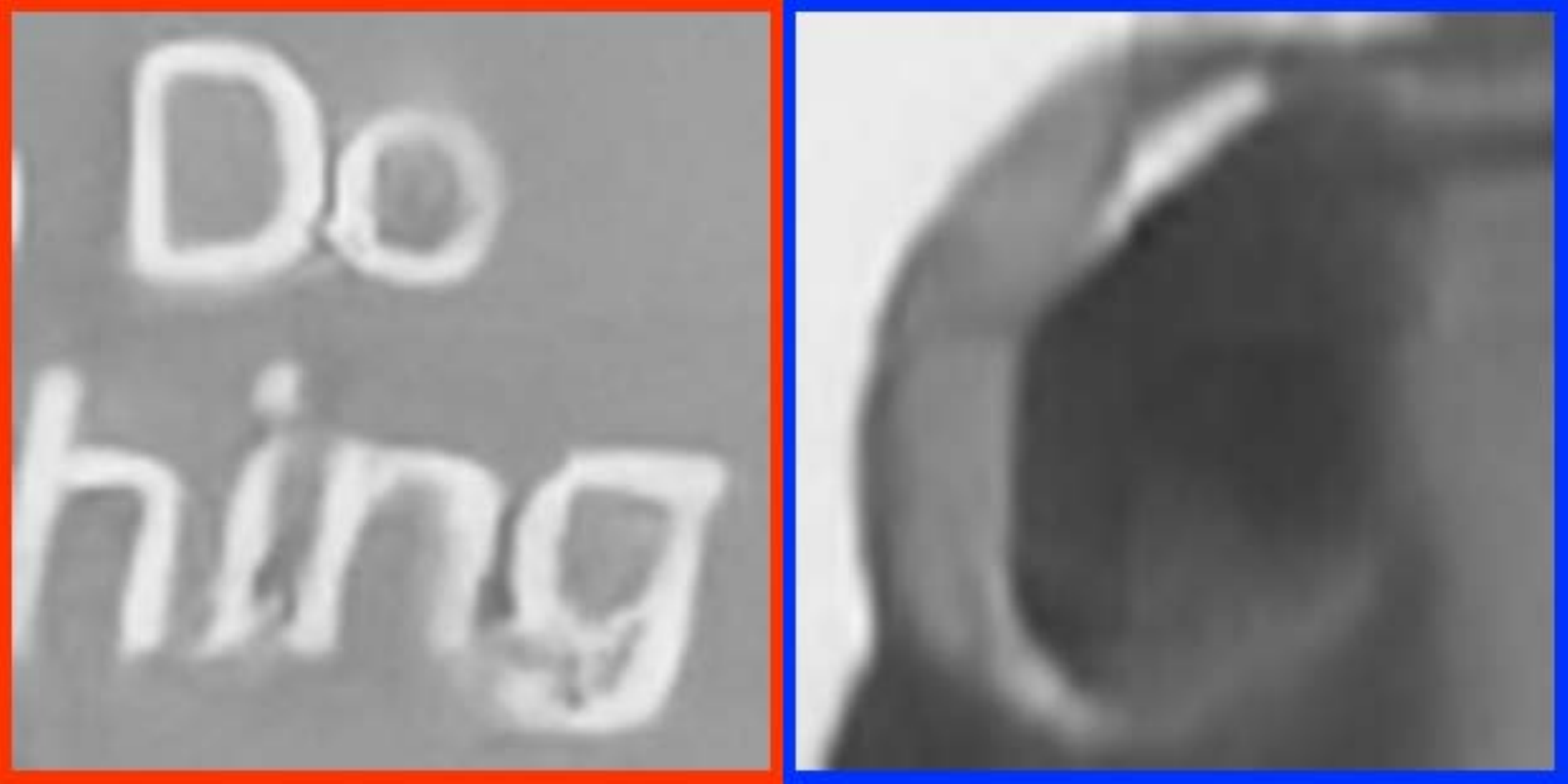} &
				\includegraphics[width=\swfive]{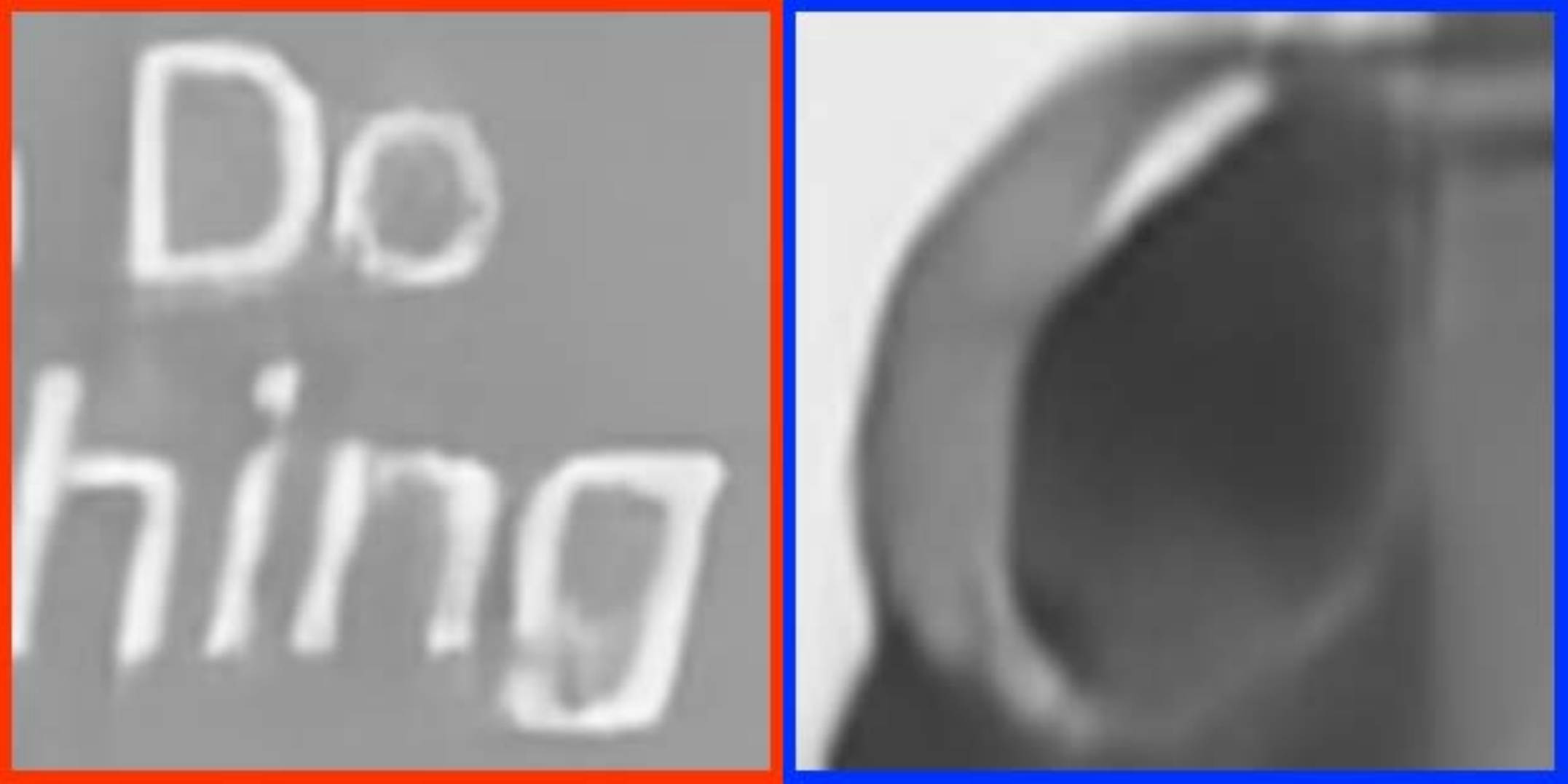} \\
				(a) clean & (b) noisy & (c) DnCNN
				%~\cite{zhang2017beyond}
				& (d) MemNet
				%~\cite{tai2017memnet} 
				& (e) SGN
				%~\cite{gu2019self}
				\\
				PSNR/SSIM & 15.22/0.354 & 23.74/0.743 & 23.78/0.742 & 23.89/0.743 \\
				\includegraphics[width=\swfive]{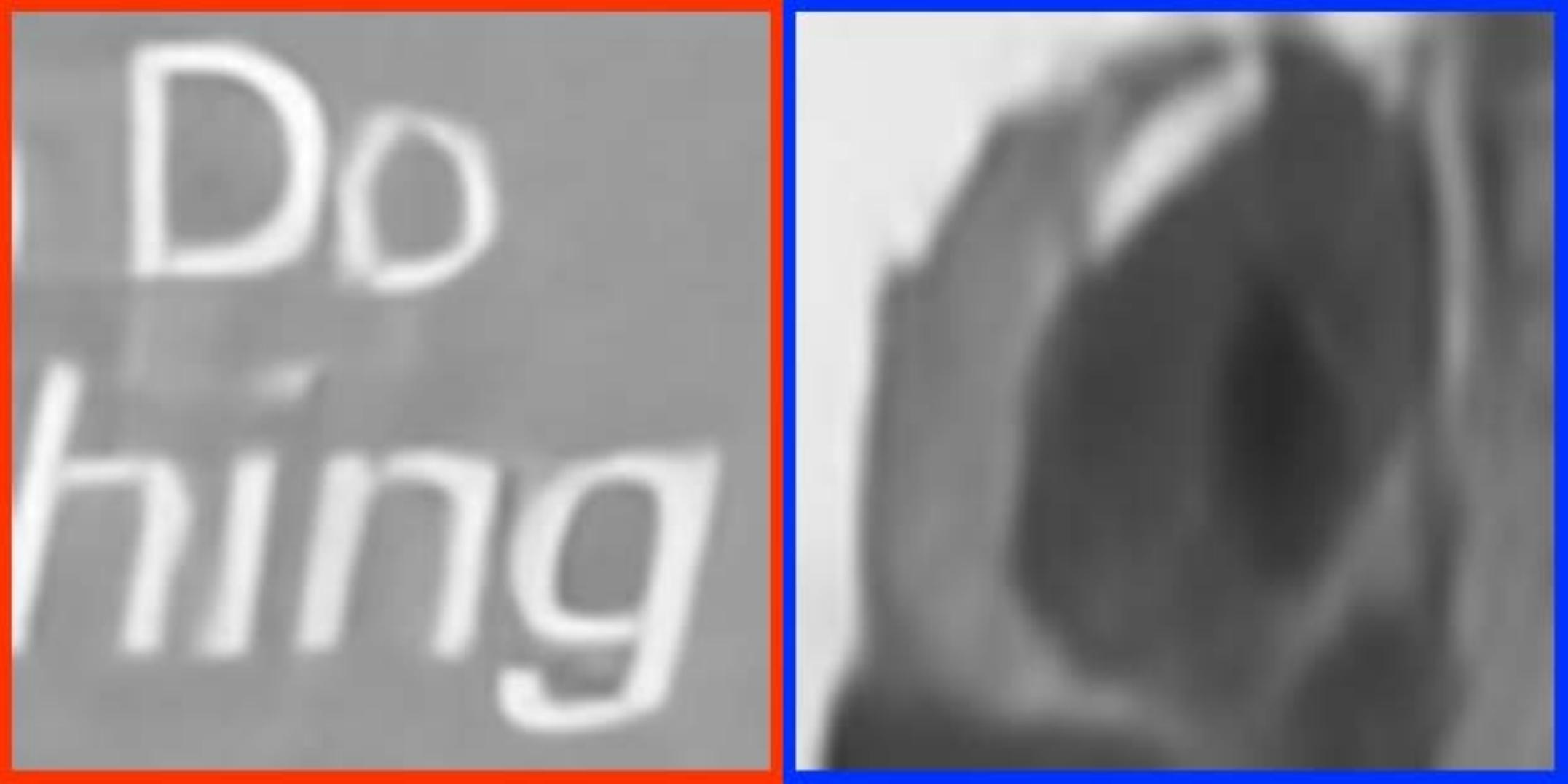} &
				\includegraphics[width=\swfive]{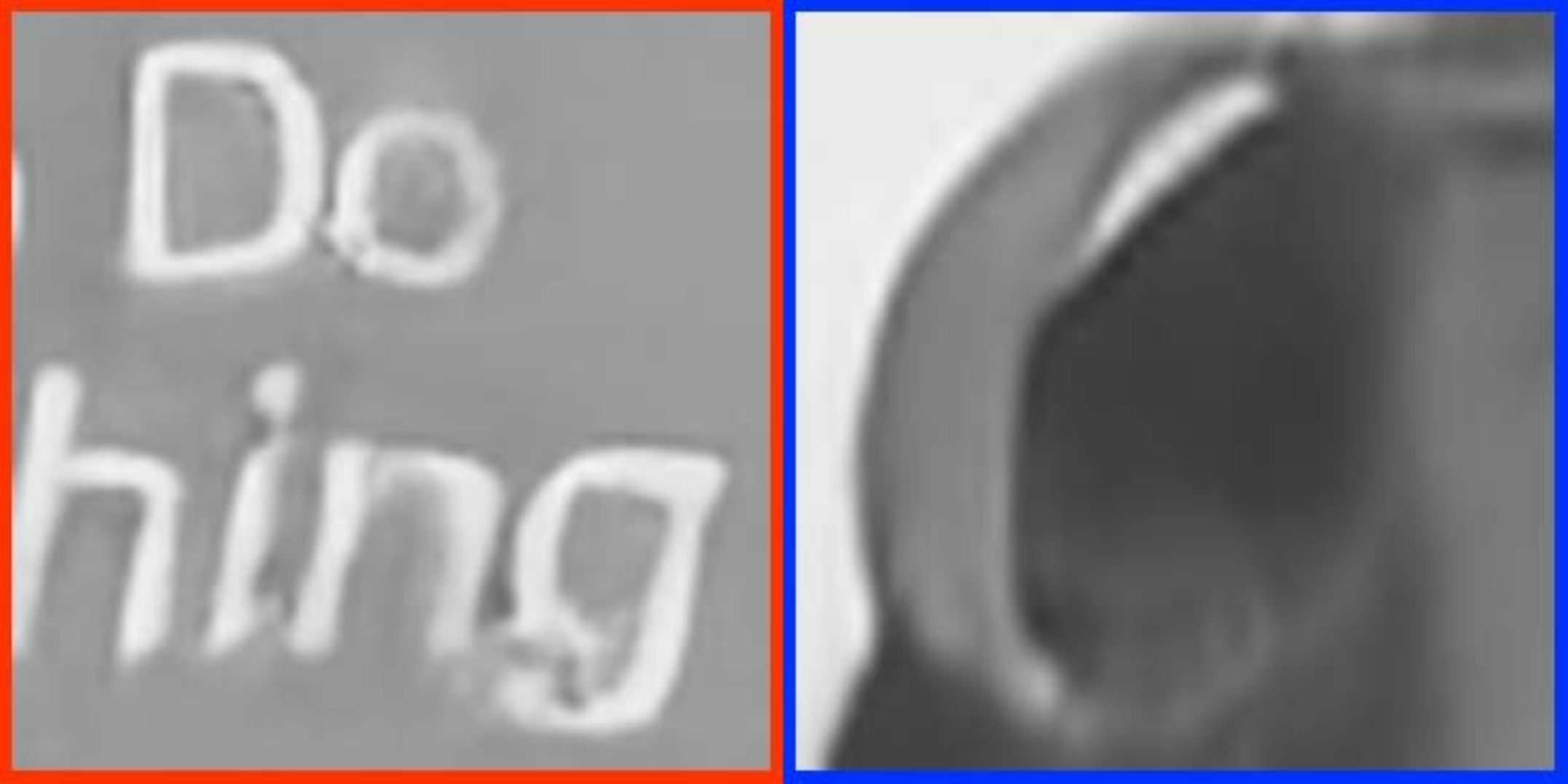} &
				\includegraphics[width=\swfive]{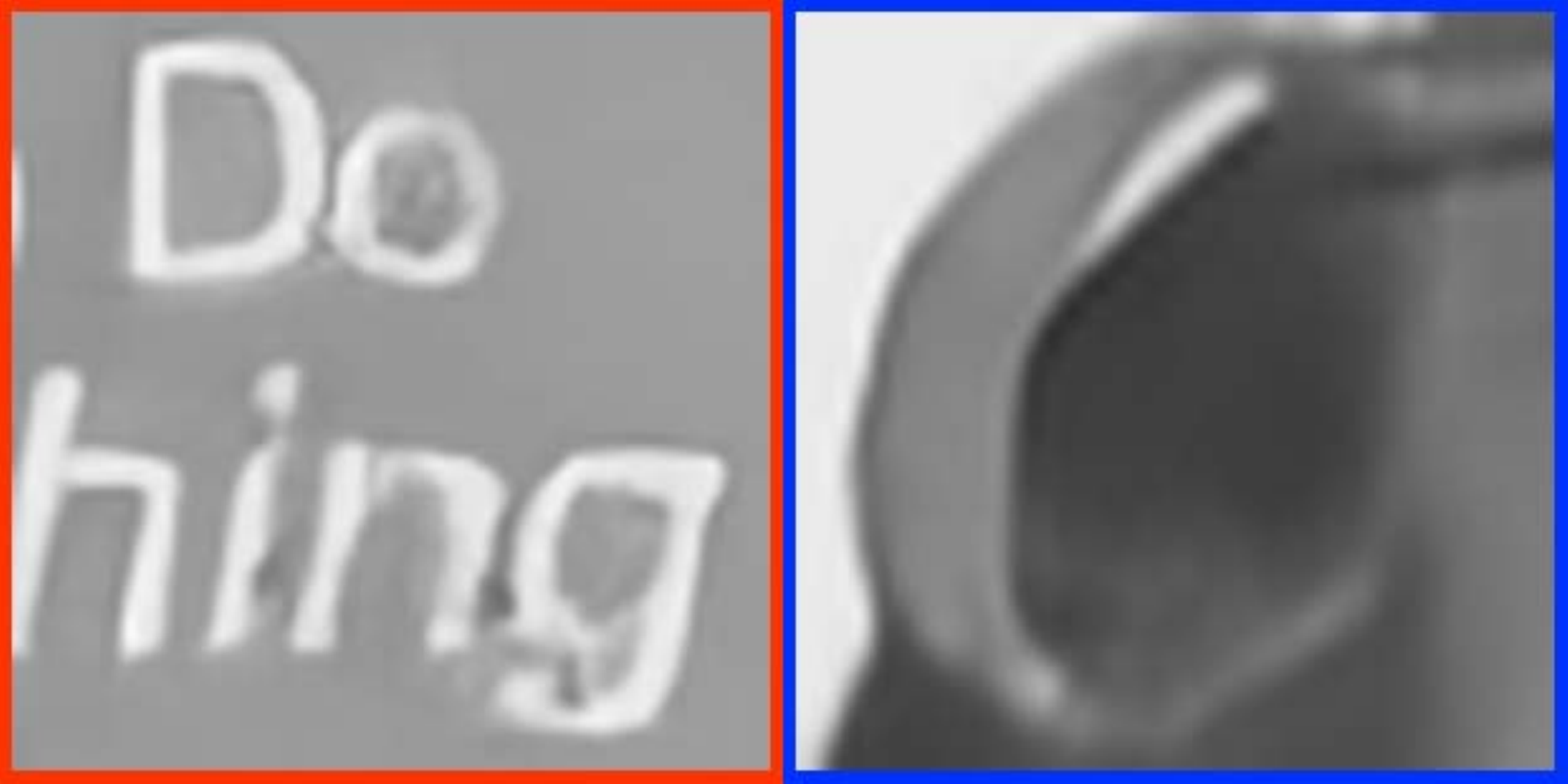} &
				\includegraphics[width=\swfive]{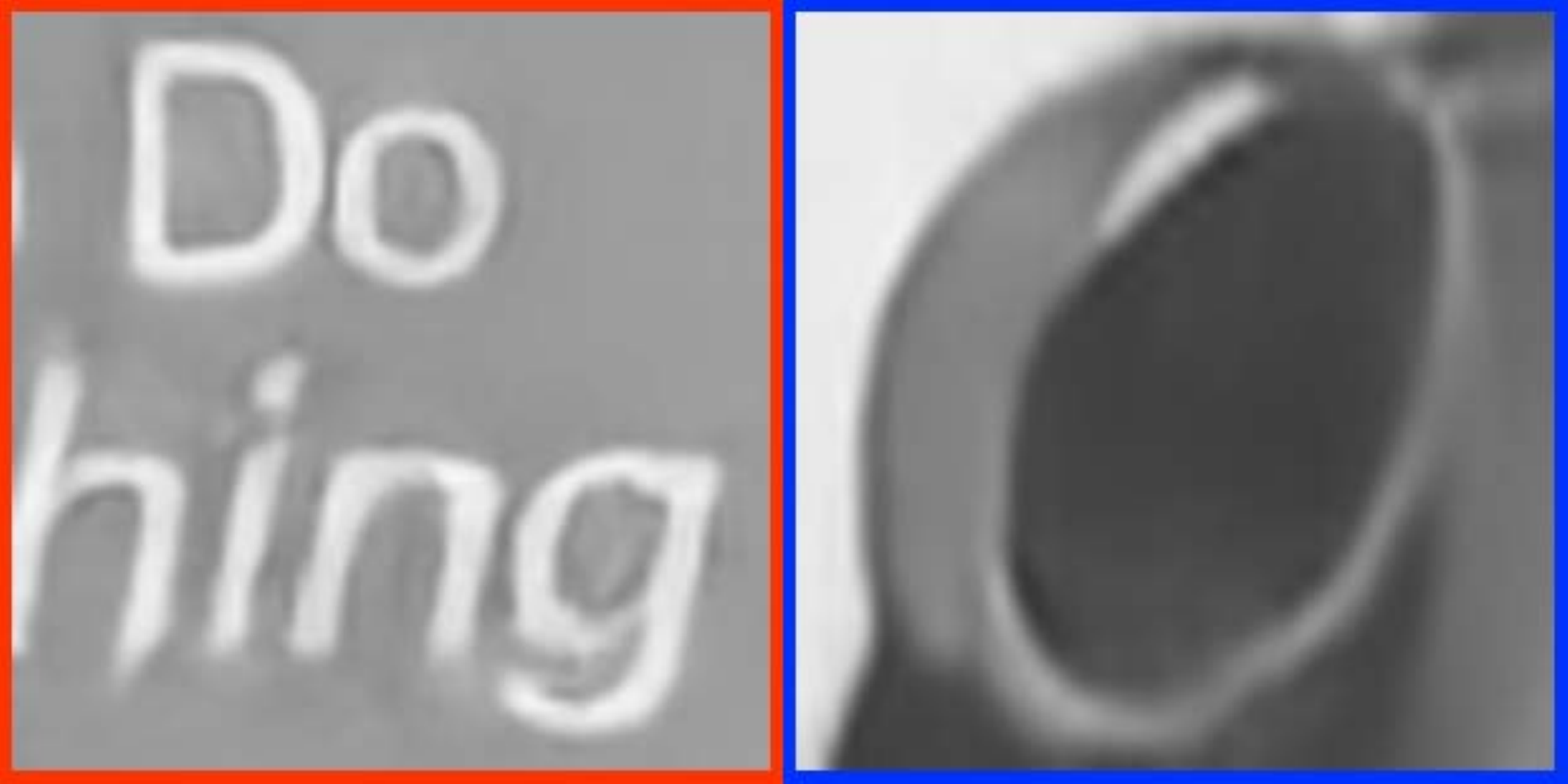} &
				\includegraphics[width=\swfive]{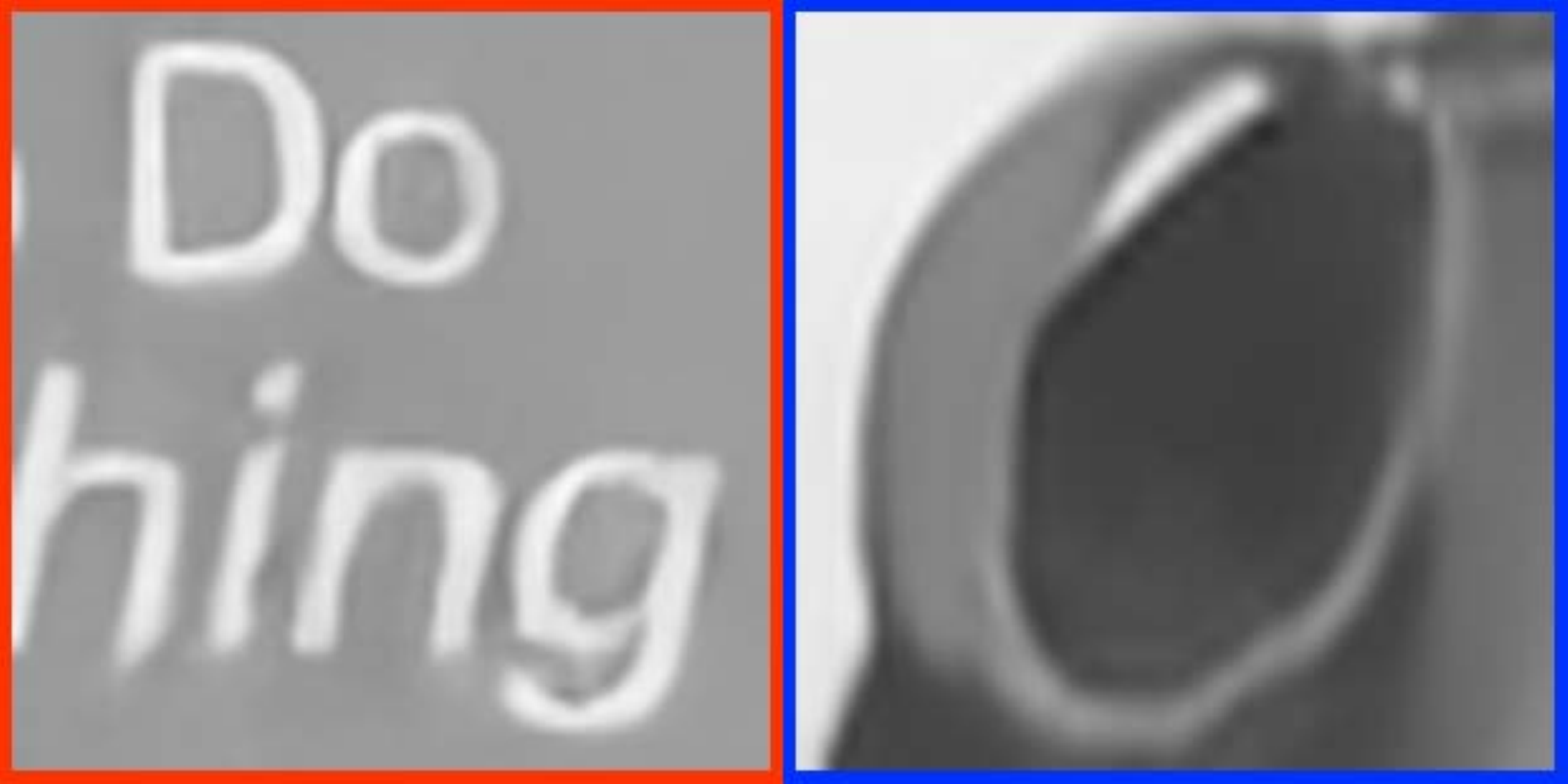} \\
				(f) FOCNet
				%~\cite{jia2019focnet}
				& (g) CARN64
				%~\cite{ahn2018fast} 
				& (h) EDSR64
				%~\cite{lim2017enhanced} 
				& (i) CS-CARN & (j) CS-EDSR \\
				24.04/0.749 & 23.87/0.739 & 24.00/0.747 & 24.37/0.771 & 24.46/0.778 \\
			\end{tabular}
		\end{center}
		%\vspace{-3mm}
		\caption{
			Visual comparison by different networks on image ``ppt3" from Set12 dataset with $\sigma =50$ noises.
			The proposed networks CS-CARN and CS-EDSR can better remove noises while preserving the correct shape of the letters and the details of the ring relative to other networks.
			%
			%Please see supplemental material for more comparisons.
		}
		\label{fig:comparisons}
	\end{figure*}
	
	\subsection{Pixel-wise Classification Network (PCN)}  \label{sec: pcn}
	We follow the idea from RAISR~\cite{romano2016raisr} which utilizes the local gradient statistics for pixel-wise classification.
	RAISR estimates the gradient orientation $\phi$, strength $\lambda$ and coherence $\mu$ for every pixel from the input noisy image and then classifies the pixels into different buckets according to a hash table
	~\footnote{Please refer to RAISR for more details to estimate $\phi$, $\lambda$, $\mu$ and hashing.}.
	Even though it is effective, this classification is still not accurate enough especially when the input is too noisy (see \reffig{fig:effective}(g)).
	As we can get the ground truth clean image during training, we can estimate a more accurate $\phi$, $\lambda$ and $\mu$.
	Then we propose a Pixel-wise Classification Network (PCN) to predict $\phi$, $\lambda$ and $\mu$ constrained by the estimated one from the ground truth clean image (see \reffig{fig:effective}(f)).
	%
	
	%\textcolor{red}{we need an example to show a image with %classes from gt, classes from noise image and our estimated %classes}.
	%
	PCN uses an architecture similar to U-Net, containing skip connections, average downsampling and bilinear upsampling to utilize multi-scale features and enlarge receptive field.
	The 3-channels output of PCN is set to the regression objective towards $\phi$, $\lambda$ and $\mu$.
	Instead of using the convolution layer with $3\times3$ filters and filter channels $c_{f}$ in each block, we adopt a Group Convolution Block (GCB) to save the computational cost.
	In GCB, each 3x3 convolution layer is replaced by a 2-layers pair as in MobileNet~\cite{howard2017mobilenets}: a $3\times3$ group convolution layer with group number 2 and output channels $\frac{c_{f}}{2}$, followed by an $1\times1$ convolution layer with output channels $c_{f}$.
	The ReLU nonlinearity is applied after each 2-layers pair.
	At the smallest scale, Group Residual Blocks, which are GCBs with residual connections, are applied after the GCB to extend the network capacity.
	%
	%	The architecture of PCN is also given in \reffig{fig:pcn}.
	%
	%Detailed configurations of PCN can be found in the supplemental material.

	%\setlength{\tabcolsep}{4pt}
	\begin{table*}[!t]
		\begin{center}
			\resizebox{\textwidth}{!}{
				\begin{tabular}{ll|lllllllllll} \hline\hline
					Dataset                   & $\sigma$                & DnCNN & CARN16 & EDSR16 & CARN64  & EDSR64  & RED   & MemNet & FOCNet & SGN     & CS-CARN & CS-EDSR \\\hline
					BSD68    &  15 & 31.97/0.878 & 31.88/0.882 & 31.99/0.887      & 32.10/0.888 & 32.16/0.887   & 32.07/0.889    & 32.14/0.891      & \textbf{32.20}/\textbf{0.893}  & 32.02/0.889          & 32.01/0.885   & 32.08/0.887   \\
					& 25 & 29.58/0.811 & 29.36/0.812 & 29.61/0.822 & 29.63/0.818 & 29.70/0.822    & 29.66/0.824 & 29.67/0.825  & 29.72/0.819  & 29.65/0.824      & 29.65/0.820   & \textbf{29.76}/\textbf{0.826}   \\
					& 50 & 26.54/0.700 & 26.55/0.700 & 26.77/0.714 & 26.70/0.707 & 26.84/0.718  & 26.73/0.712 & 26.74/0.709  & 27.04/0.720  & 26.77/0.710    & 27.21/0.731   & \textbf{27.31}/\textbf{0.738}   \\ \hline
					Urban100 & 15   & 32.85/0.920 & 32.21/0.917 & 32.75/0.921   & 32.95/0.920 & 33.10/0.931  & 32.59/0.923     & 32.50/0.919      & \textbf{33.15}/\textbf{0.927}  & 32.57/0.911         & 32.76/0.921   & 32.93/0.925   \\
					& 25   & 30.04/0.869  & 29.27/0.862 & 29.96/0.878  & 30.02/0.879 & 30.36/0.883   & 29.93/0.880 & 29.91/0.878  & \textbf{30.64}/\textbf{0.887}  & 30.05/0.880    & 29.83/0.874 & 30.17/0.882 \\
					& 50   & 26.30/0.772  & 25.77/0.755 & 26.44/0.781  & 26.33/0.774 & 26.85/0.794  & 26.29/0.779     & 26.34/0.778      & \textbf{27.10}/0.803  &  26.35/0.781          & 26.76/0.799   & 27.07/\textbf{0.810}   \\ \hline
					DIV2K    & 15   & 33.93/0.901 & 33.75/0.901 & 33.93/0.905    & 34.04/0.906 & \textbf{34.14}/\textbf{0.908}  & 34.04/0.907     & 33.99/0.905      & -      & 34.02/0.902     & 33.95/0.904   & 34.03/0.906   \\
					& 25   & 31.39/0.845 & 31.16/0.846 & 31.52/0.856 & 31.53/0.856 & 31.66/0.855  & 31.56/0.859 & 31.58/0.858  & -      & 31.68/0.860      & 31.58/0.856   & \textbf{31.74}/\textbf{0.861}   \\
					& 50   & 28.25/0.753 & 28.18/0.751 & 28.51/0.765 & 28.40/0.759 & 28.65/0.766  & 28.53/0.769     & 28.43/0.753      & -      & 28.53/0.771     & 29.02/0.782   & \textbf{29.17}/\textbf{0.789}   \\ \hline
					Set12    & 15   & 33.16/0.887 & 32.90/0.893 & 33.11/0.897    & 33.16/0.896 & 33.34/\textbf{0.899}  & 33.11/0.891     & 33.15/0.897      & \textbf{33.37}/0.896  & 33.07/0.881       & 33.14/0.896   & 33.18/0.898   \\
					& 25   & 30.73/0.840 & 30.40/0.837 & 30.78/0.849    & 30.76/0.845 & 30.95/0.850  & 30.96/0.851 & 30.78/0.850  & 30.73/0.846  & 30.81/0.850      & 30.78/0.847   & \textbf{30.97}/\textbf{0.853}   \\
					& 50   & 27.44/0.747  & 27.28/0.737 & 27.64/0.755  & 27.58/0.748 & 27.86/0.762  & 27.50/0.766 & 27.48/0.749  & 27.99/0.765  & 27.62/0.758       & 28.16/0.773   & \textbf{28.33}/\textbf{0.781}  \\ \hline
					\multicolumn{2}{c|}{kFLOPs/pixel}        & 1108.2 & 72.6 & 145.1 & 702.7  & 2361.6  & 1106.4   & 5480.0 & 2225.7 & 429.1   & 75.5 & 148.0 \\ \hline \hline
			\end{tabular}}
		\caption{Denoising performance (PSNR/SSIM) for $\sigma = 15,25,50$ noises on BSD68, Urban100, DIV2K, Set12, and computational cost (kFLOPs/pixel) by different methods. For CS-CARN and CS-EDSR, the numbers are slightly higher than CARN16 and EDSR16 because the additional computational cost of the PCN model (2.8 kFLOPs/pixel) is included. The symbol ``-" denotes that the corresponding results are not provided.}
		\label{table:compare}
		\end{center}
	\end{table*}

	\subsection{Class Specific Convolution (CSConv)}  \label{sec: csconv}
	Different from KPN which uses a large network to estimate the pixel-wise kernels, we use the above efficient PCN to classify pixels into different classes and then learn different weights for different classes in the proposed class specific convolution (CSConv).
	Specifically, the $i$-th learnable weights $W_i$ of CSConv will be fetched from the filter bank $W$ and used to filter the input feature $Q(m,n,c)$ if the pixel at position $(m,n)$ is classified as the $i$-th class from the noisy image:
	\begin{eqnarray}
	\label{eq:csconv}
	\hat{Q}(m,n,c) = \sum_{s=-r}^r\sum_{t=-r}^r\sum_{c'=0}^{C_{in}-1}[ Q(m-s,n-t,c')\nonumber\\\times W_i(s,t,c,c')]
	%\vspace{-1mm}
	\end{eqnarray}
	in which $\hat{Q}$ is the output feature and $K=2r+1$ is the kernel size.
	$s$ and $t$ are the index of the kernel in two coordinates and also the offset relative to the center pixel $(m, n)$ in the feature map to be filtered by the kernel.
	By using CSConv, we actually use different weights to deal with different classes of pixels.
	During training, the training loss will be passed to each chosen filter $W_{i}$ through back propagation and then be updated by the optimizer.
	%
	%	Even though the input and output channels are reduced which will lead to fewer computational costs, there still exist abundant weights to support the proposed network to remove noises from different contents.
	Even though the input and output channels are reduced to save the computational costs, the proposed network still contains enough capacity to remove noises from different contents.

	\subsection{CSConv-based Denoising Convolutional Network (CSDN)}  \label{sec: csdn}
	In this work, we integrate the proposed CSConv into the EDSR and CARN architecture as the proposed CSDN in our experiments\footnote{CSConv can also be easily integrated into other existing denoising networks by replacing any convolution layer.}.  
	The baseline model EDSR consists of a 1-layer source encoder, a feature extractor containing 16 residual blocks in series, and a 1-layer output decoder.
	A residual connection is applied after the last residual block.
	In CARN, the feature extractor contains three cascading blocks.
	Each cascading block has three efficient residual blocks that utilize group convolution and pointwise convolution, and adds cascading connections to merge features from the residual blocks.
	Global cascading connections similar to the ones inside cascading blocks are added to merge cascading block outputs.  
	%
	%	Unlike RAISR\cite{romano2016raisr} and KPN\cite{mildenhall2018burst} that apply dedicated kernels at the final linear filtering step, CSConv can replace the convolution layers in EDSR and CARN.
	Unlike RAISR and KPN that directly apply dedicated kernels to the input images, CSConv can replace the convolution layers in EDSR and CARN.
	Specifically, the second convolution layers in each residual block of both EDSR and CARN are replaced by CSConv in our experiments. 
	All filters are in size $3\times 3$, and PReLU nonlinearity~\cite{he2015delving} is applied between two convolution layers in each residual block.
	To make the proposed CSDN more efficient, we reduce the filter channels by a factor of four relative to the original EDSR and CARN.
	%
	%Detailed configurations of CS-EDSR and CS-CARN can be found in the supplemental material.
	
		\begin{figure*}[!t]
		\begin{center}
			\begin{tabular}{cccccc}
				%				\includegraphics[width=\swsix]{figs3/img079_x1_HR.pdf} &
				%				\includegraphics[width=\swsix]{figs3/img079_x1_LR.pdf} &
				%				\includegraphics[width=\swsix]{figs3/img079_x1_SR_csedsr.pdf} &
				%				\includegraphics[width=\swsix]{figs3/img079_x1_SR_kpnplus.pdf} &
				%				\includegraphics[width=\swsix]{figs3/img079_x1_SR_kpn.pdf}&
				%				\includegraphics[width=\swsix]{figs3/img079_raisr.pdf}
				%				\\
				\includegraphics[width=\swsix]{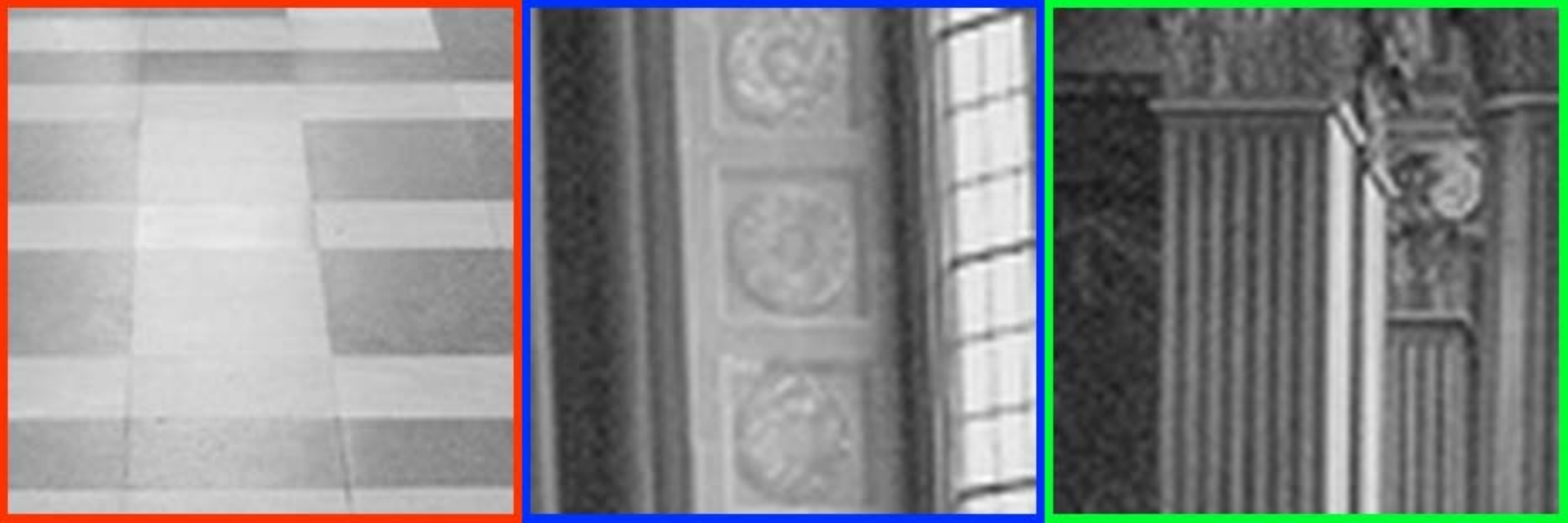} &
				\includegraphics[width=\swsix]{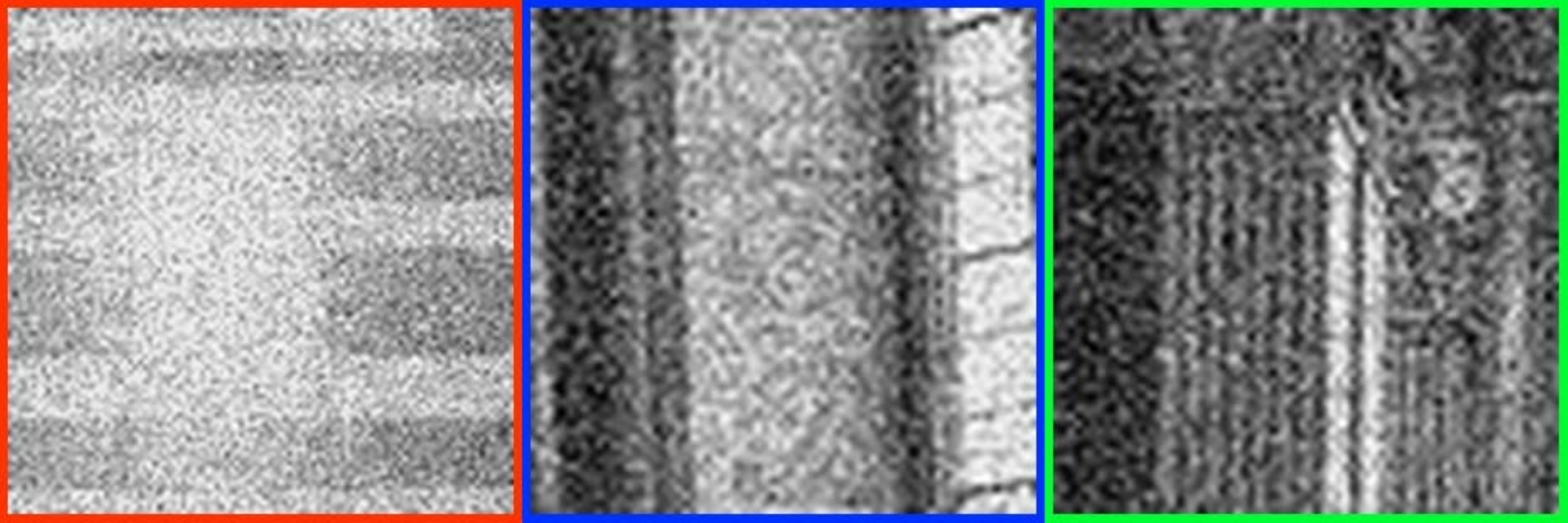} &
				\includegraphics[width=\swsix]{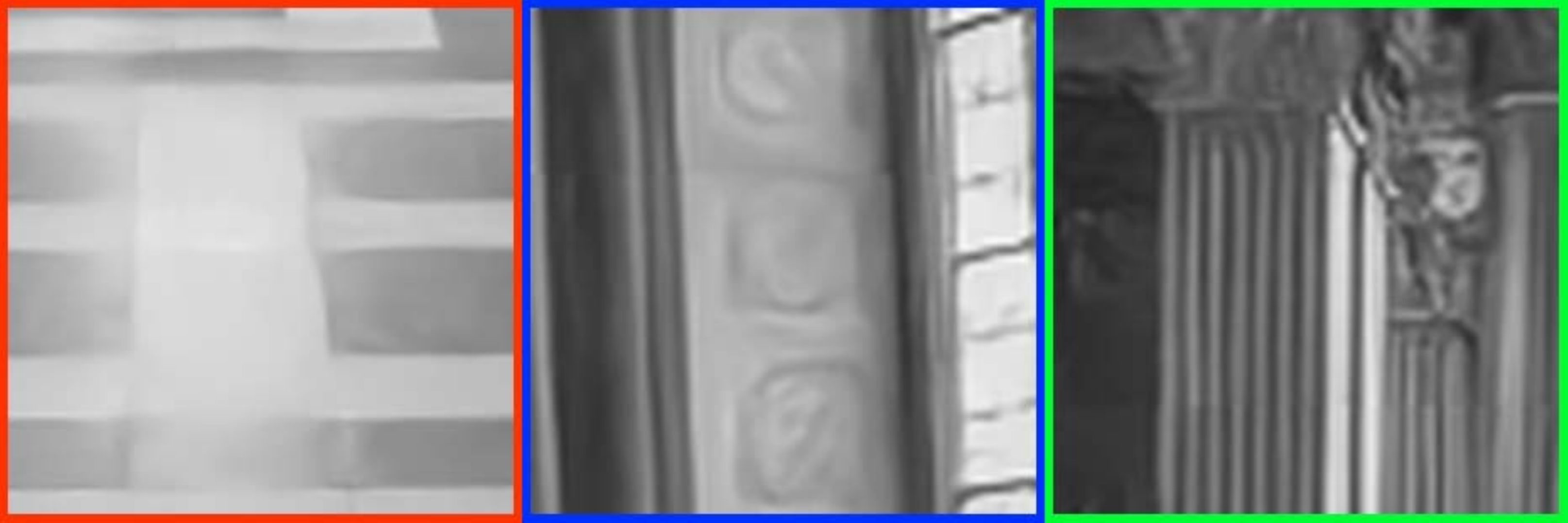} &
				\includegraphics[width=\swsix]{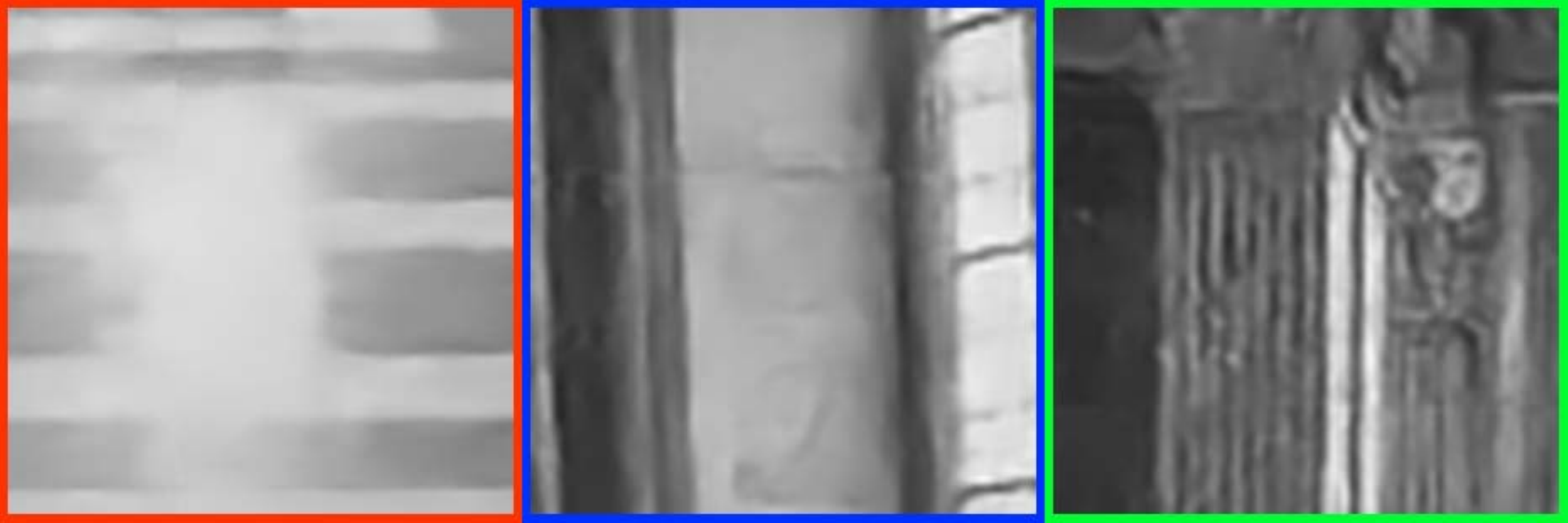} &
				\includegraphics[width=\swsix]{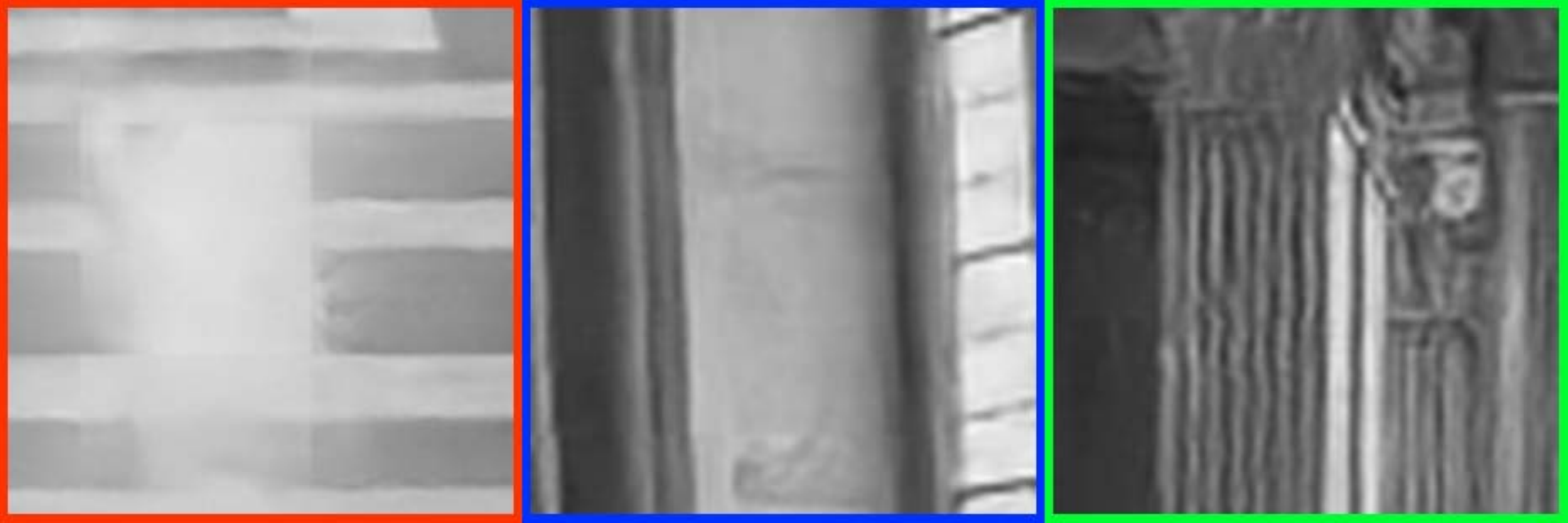}&
				\includegraphics[width=\swsix]{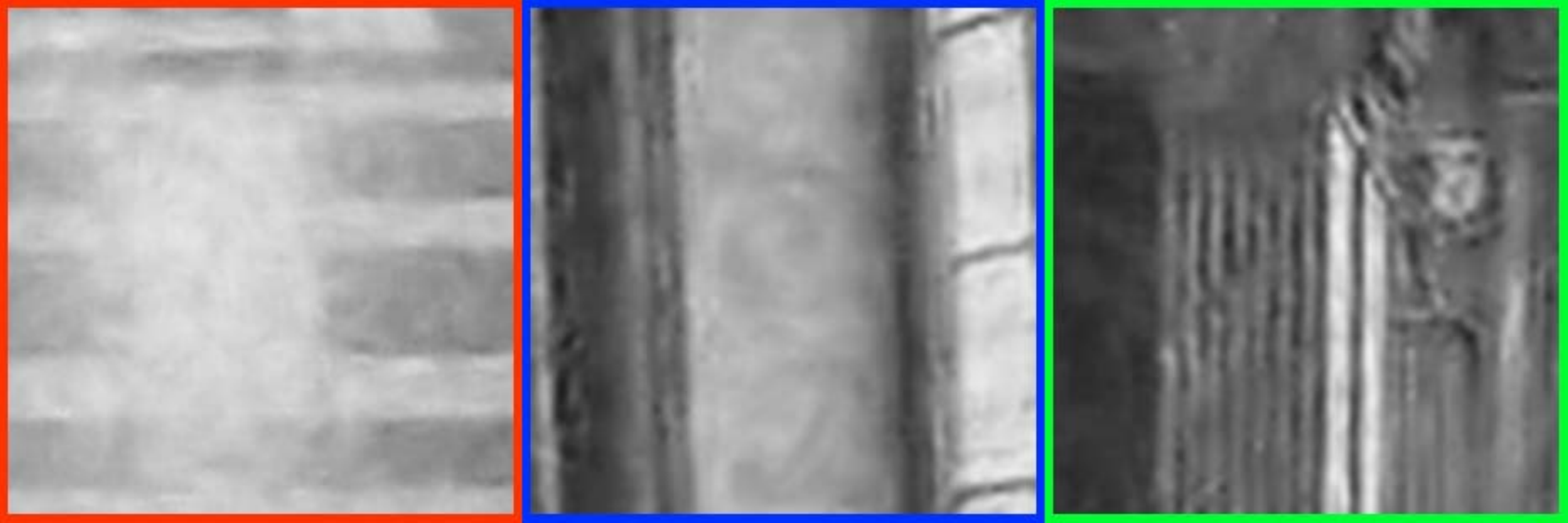} \\
				(a) clean & (b) noisy & (c) CS-EDSR & (d) KPN+ & (e) KPN
				%\cite{mildenhall2018burst}
				& (f) RAISR
				%\cite{romano2016raisr}
				\\
				PSNR/SSIM & 21.16/0.596 & 26.86/0.869  & 26.19/0.843 & 26.47/0.855 & 25.42/0.800  \\
			\end{tabular}
		\end{center}
		%\vspace{-3mm}
		\caption{
			Visual comparison among CS-EDSR, KPN+, KPN and RAISR on image ``img079" from Urban100 dataset with $\sigma =25$ noises.
			In the zoomed area, CS-EDSR better recovers the textures and details compared to other methods. 
			KPN and KPN+ show stronger over-smoothing effects and fail to recover edges. RAISR has artifacts and remaining noises.
		}
		\label{fig:raisrkpn}
	\end{figure*}
	
	\subsection{Loss Functions}  \label{sec: loss}
	The loss functions of PCN and CSDN depend on the pixel-wise class as well as the clean image.
	For PCN, we treat $\phi$, $\lambda$ and $\mu$, which are the gradient statistics from RAISR estimated from the clean image, as the ground truth of $\hat{\phi}$, $\hat{\lambda}$ and $\hat{\mu}$ and optimize the parameters $\theta_C$.
	Then, the loss function of PCN can be written as follows:
	\begin{eqnarray}
	\label{eq:pcnloss}
	\mathcal{L}^{PCN}(Y;\theta_C) &=& ||\hat{\phi}(Y;\theta_C) - \phi||_1 + ||\hat{\lambda}(Y;\theta_C) - \lambda||_1 \nonumber\\&+& ||\hat{\mu}(Y;\theta_C) - \mu||_1,
	%\vspace{-1mm}
	\end{eqnarray}

	in which $||\cdot||_1$ is the $L_1$ norm.
	After using the hash table in RAISR, $\hat{\phi}$, $\hat{\lambda}$ and $\hat{\mu}$ can further be quantized into different classes by $H=\mathbf{H}(\hat{\phi}, \hat{\lambda}, \hat{\mu})$ as we mentioned above.
	As to CSDN, we also use the $L_1$ loss to minimize the difference between the estimated image $\hat{X}$ and ground truth clean image $X$ by optimizing the parameters of CSDN $\theta_D$ as:
	\begin{equation}
	\label{eq:csdnloss}
	\mathcal{L}^{CSDN}(Y,H;\theta_D) = ||\hat{X}(Y,H;\theta_D) - X||_1.
	%\vspace{-1mm}
	\end{equation}

	\setlength{\tabcolsep}{1pt}
	\begin{table}[!t]
		\begin{center}
			
			%\resizebox{\linewidth}{!}{
			\begin{tabular}{l|llll} \hline \hline
				%Dataset  & CS & 16 & G & N & $\phi_4$ & CE & CS & 16 & G & N & $\phi_4$ & CE \\ \hline
				
				Dataset   & CS-EDSR   & KPN+ & KPN  & RAISR  \\ \hline
				BSD68       & 29.76/0.826   & 29.18/0.807 & 29.55/0.822 & 28.15/0.754    \\
				Urban100    & 30.17/0.882   & 28.62/0.851 & 29.72/0.875 & 27.56/0.793     \\
				DIV2K     & 31.74/0.861   & 31.02/0.842 & 31.51/0.857 & 29.80/0.788     \\
				Set12       & 30.97/0.853   & 29.99/0.835 & 30.67/0.848 & 28.76/0.782   \\ \hline\hline
			\end{tabular}
			%}
			\caption{Denoising performance (PSNR/SSIM) by CS-EDSR, KPN+, KPN and RAISR for $\sigma = 25$ noises.}
			\label{table:raisr}
		\end{center}
	\end{table}

	\section{Experiments and Results}
	\subsection{Datasets}
	Our training set consists of 400 images from BSD500~\cite{martin2001database}, 800 images from DIV2K~\cite{agustsson2017ntire}, 4744 images from Waterloo~\cite{ma2016waterloo}, and 5000 images from 5K~\cite{bychkovsky2011learning}. 
	The same set of training images are used to train both PCN and CSDN.
	BSD68\cite{martin2001database}, Set12\cite{zhang2017beyond}, Urban100\cite{huang2015single}, and 10 images from DIV2K\cite{agustsson2017ntire} are used for evaluation.
	For both training and test sets, we generate noisy images with additive white Gaussian noises (AWGN) with standard variation $\sigma=15,25,50$.

	\subsection{Experimental Setting}
	The proposed model and experiments are implemented with the PyTorch library.
	For the proposed PCN and CSDN, we set the numbers of classes for  $\phi$, $\lambda$ and $\mu$ as 8, 3 and 3, respectively.
	As the hashing procedure is non-differentiable, we train PCN and CSDN separately.
	We first train the PCN to ensure the estimation quality of pixel-wise classification.
	Then, we fix the parameters of PCN and train CSDN.
	When training, we choose batch size as 4 and patch size as 96.
	Data augmentation including random flip and $0^\circ$, $90^\circ$, $180^\circ$, $270^\circ$ rotation are adopted when generating the training patches.
	ADAM optimizer ~\cite{kingma2014adam} is used in training with $\beta_1 = 0.9, \beta_2 = 0.999, \epsilon = 1 \times 10^{-8}$.
	The initial learning rate is set to $10^{-4}$, and decays by factor 0.5 after every 20 epochs.
	Both PCN and CSDN are trained for 100 epochs.
	
	\subsection{Experimental Results}
	The experimental results of proposed methods are compared with the following state-of-the-art deep image denoising networks including DnCNN~\cite{zhang2017beyond}, RED~\cite{mao2016image}, MemNet~\cite{tai2017memnet}, SGN~\cite{gu2019self} and FOCNet~\cite{jia2019focnet}.  
	As to our baseline structures CARN and EDSR, the original implementations have 64 features in every residual or cascading block.
	We denote them as CARN64 and EDSR64 in the following experiments.
	The proposed network CS-CARN and CS-EDSR only contain 16 features in every block to reduce the computational cost.
	For comparison, we also train CARN and EDSR with only 16 features denoted as CARN16 and EDSR16, respectively.
	Except for SGN and FOCNet which use the publicly available implementation, we re-implement the networks by PyTorch and all the networks are trained with the same training set described above.

	\reftable{table:compare} shows the average PSNR and SSIM of different methods under three noise levels on four evaluation datasets as well as their FLOPs per pixel.
	Specifically, the proposed networks CS-CARN and CS-EDSR perform favorably over state-of-the-art methods on BSD68, DIV2K and Set12 under noise variance $\sigma =25,50$.
	It is worth noting that the performance gain of CSConv increases as the noise level becomes larger. 
	Even though CS-CARN and CS-EDSR cannot compete with some other larger networks when $\sigma=15$ or on Urban100, they are not worse than the efficient denoise FLOPs per pixel.
	In addition, CARN16 and EDSR16 perform much worse than CS-CARN and CS-EDSR with almost the same computational cost which demonstrates the effectiveness of the proposed CSConv.

	We also show some visual comparisons on Set12 dataset with $\sigma =50$ noises in \reffig{fig:comparisons}, where the proposed method recovers finer details and avoids over-smoothing in the denoised image.

	\begin{table}[!t]
		\begin{center}
			
			%\vspace{-2mm}
			
			%\resizebox{\linewidth}{!}{
			\begin{tabular}{l|lll|lll} \hline \hline
				& \multicolumn{3}{c|}{PCN}  & \multicolumn{3}{c}{RAISR+N} \\
				Dataset  & $\sigma = 15$  & $\sigma = 25$ & $\sigma = 50$  & $\sigma = 15$  & $\sigma = 25$ & $\sigma = 50$   \\ \hline
				BSD68    & 0.030        & 0.031 & 0.035   & 0.052           & 0.062 & 0.075   \\
				Urban100 & 0.025       & 0.026 & 0.025   &  0.051       & 0.068 & 0.092   \\
				DIV2K    & 0.028         & 0.030 & 0.032   & 0.059       & 0.069 & 0.081   \\
				Set12    & 0.030         & 0.031 & 0.032   & 0.052      & 0.064 & 0.080  \\ \hline\hline
			\end{tabular}
			%}
		\caption{
			Comparison of mean square error (MSE) of the estimated gradient statistics by PCN and RAISR (denote as ``RAISR+N'') from noisy images on different datasets and noise levels.
			The ground truth gradient statistics are estimated from clean images by RAISR.
			It shows that the proposed PCN can estimate more accurately from noisy images than RAISR.
		}	
	\label{table:mse}
		\end{center}
	\end{table}

	\begin{figure*}[!t]
		\begin{center}
			\begin{tabular}{ccccc}
				%				\includegraphics[width=\swfive]{figs5/0041_x1_HR.pdf}&
				%				\includegraphics[width=\swfive]{figs5/0041_x1_LR.pdf}&
				%				\includegraphics[width=\swfive]{figs5/0041_x1_SR_gt.pdf}&
				%				\includegraphics[width=\swfive]{figs5/0041_x1_SR_noise.pdf} &
				%				\includegraphics[width=\swfive]{figs5/0041_x1_SR_csedsr.pdf}
				%				\\
				\includegraphics[width=\swfive]{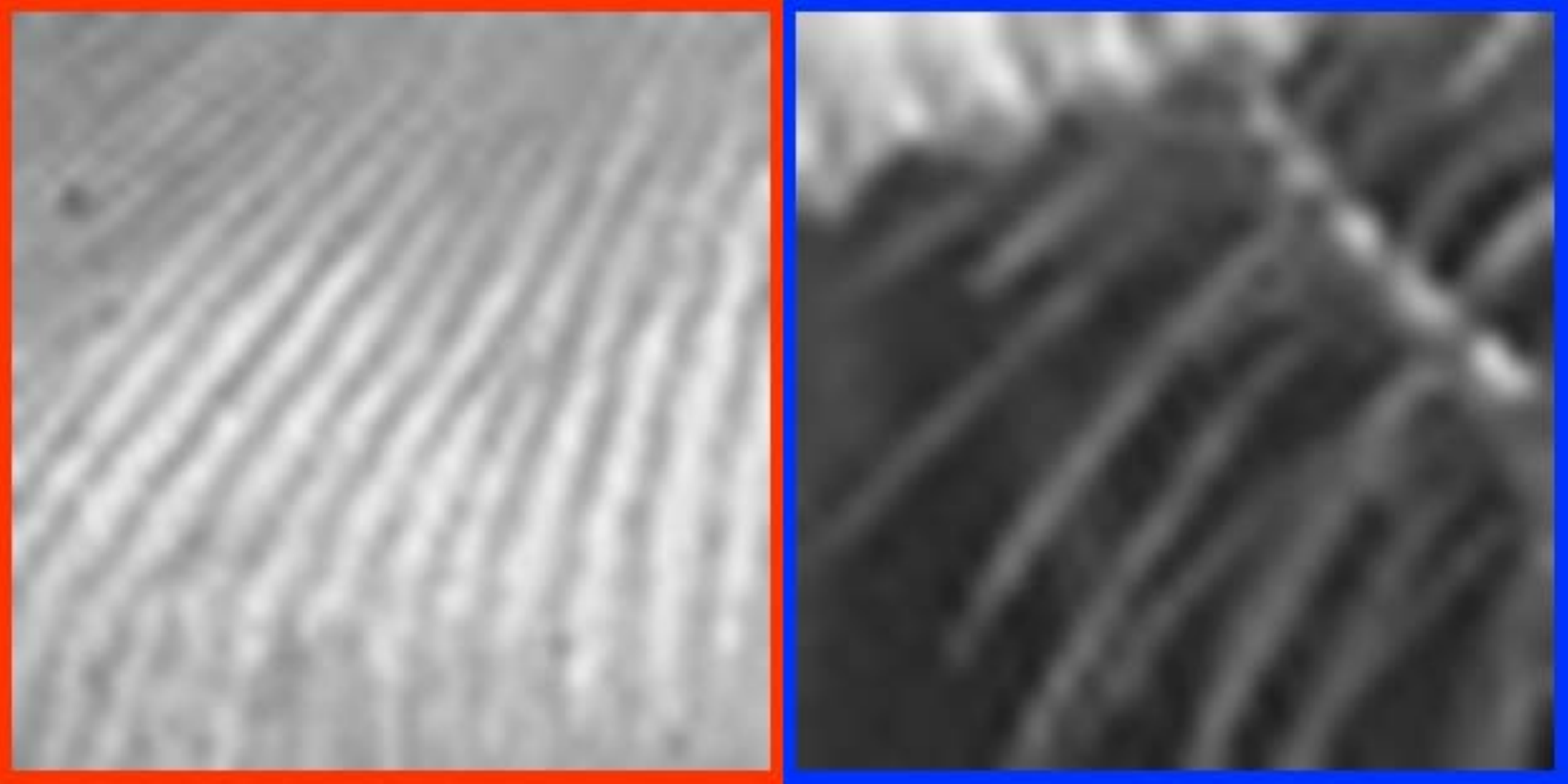}&
				\includegraphics[width=\swfive]{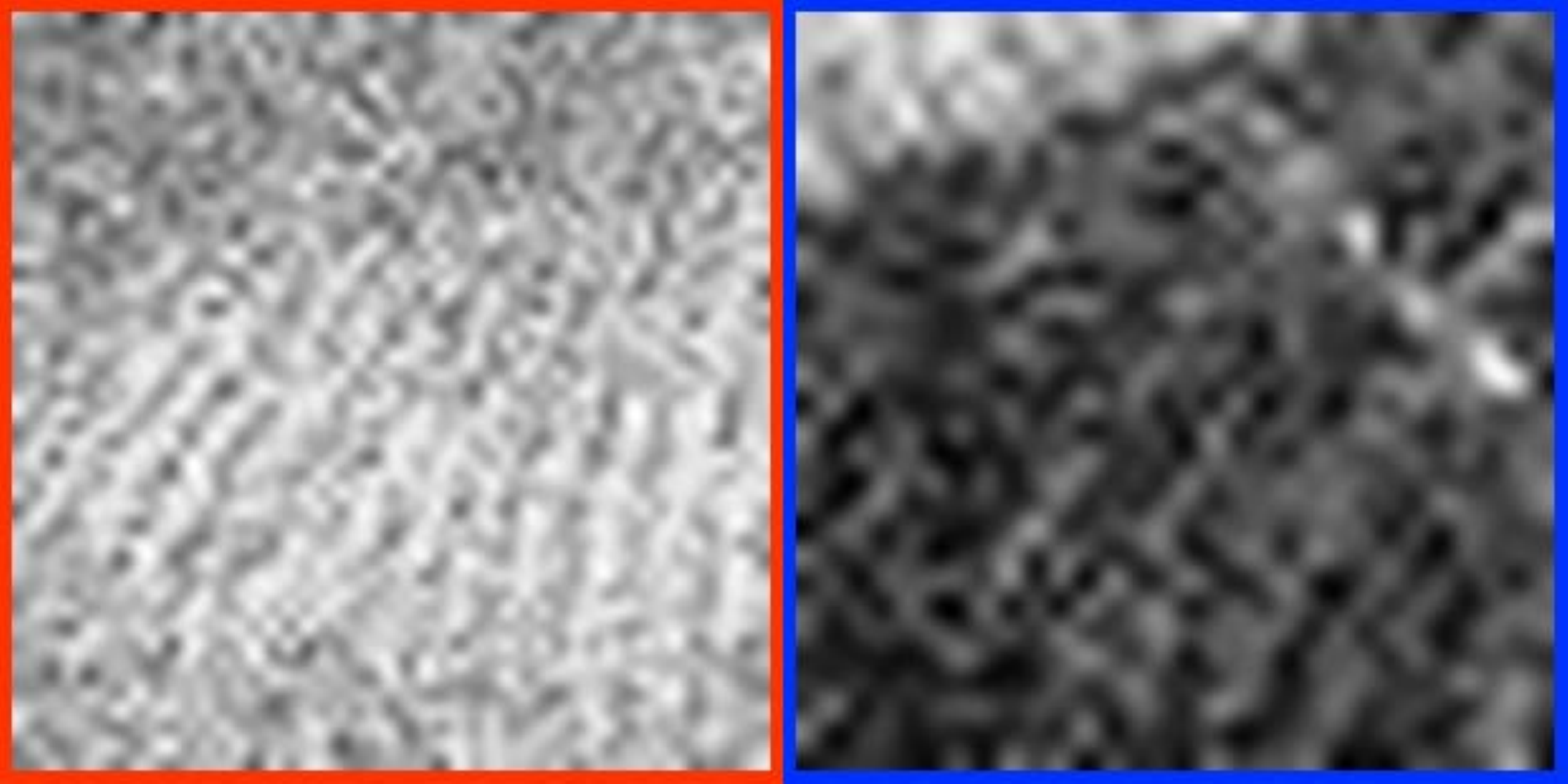}&
				\includegraphics[width=\swfive]{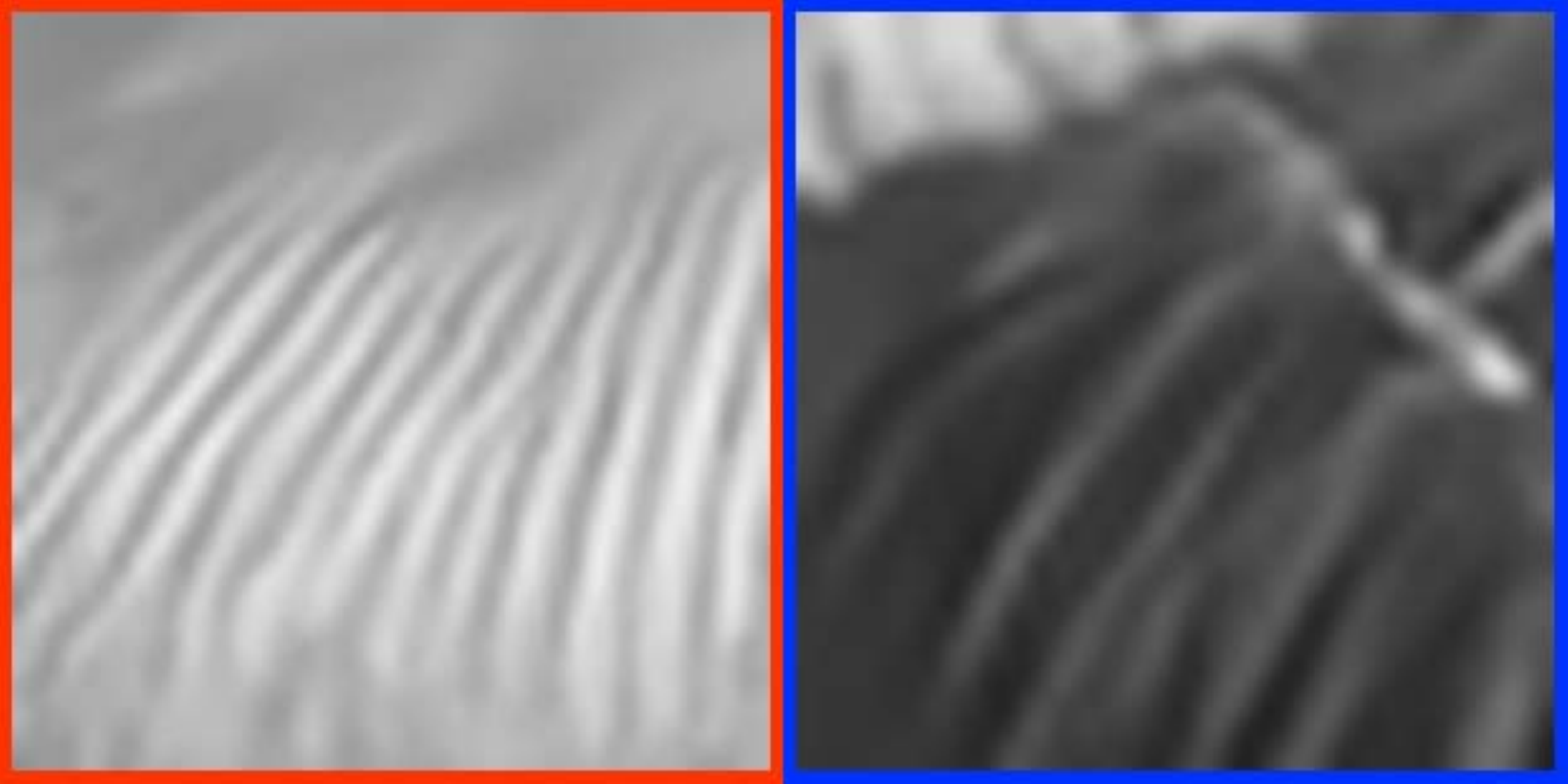}&
				\includegraphics[width=\swfive]{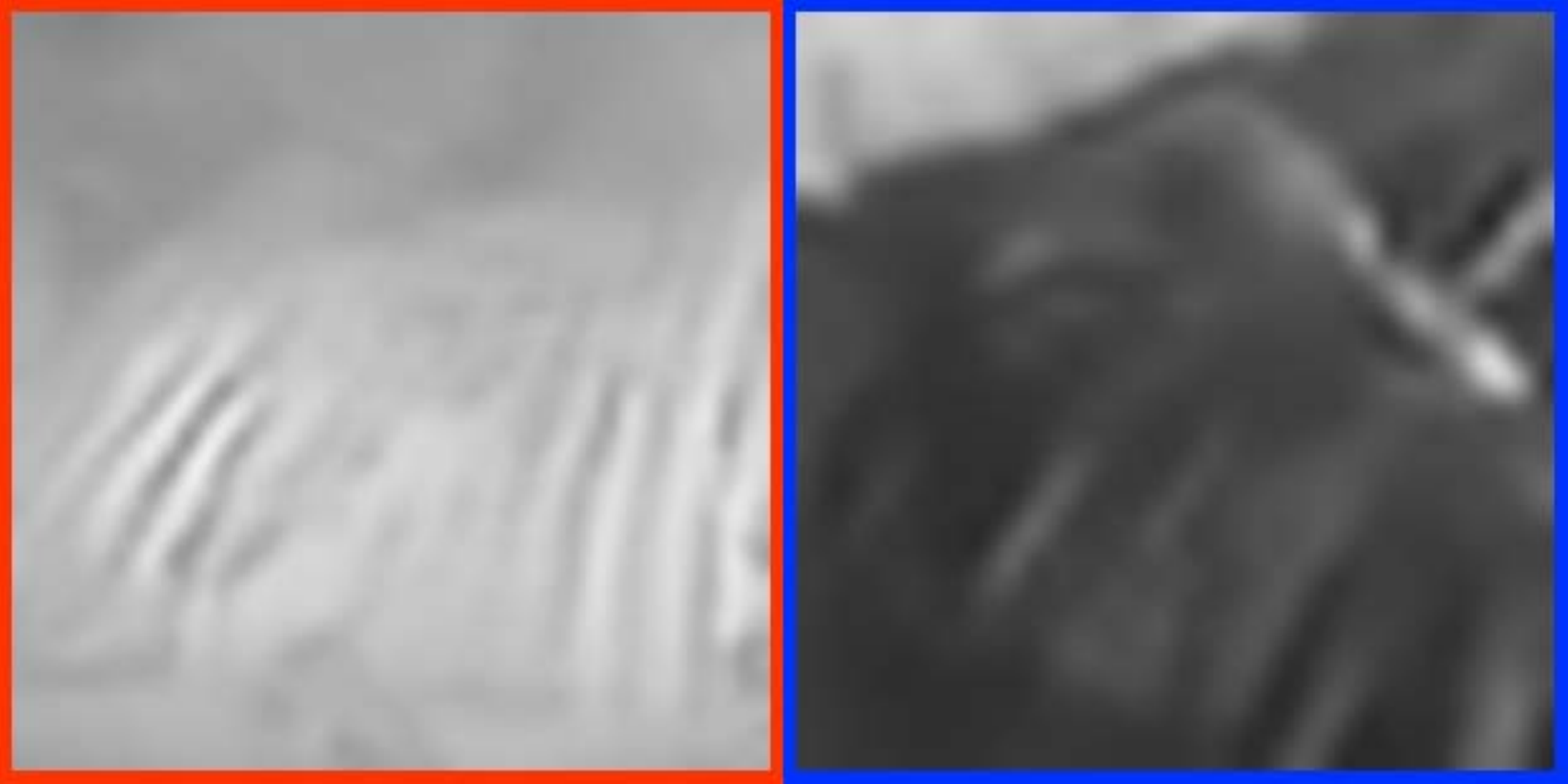} &
				\includegraphics[width=\swfive]{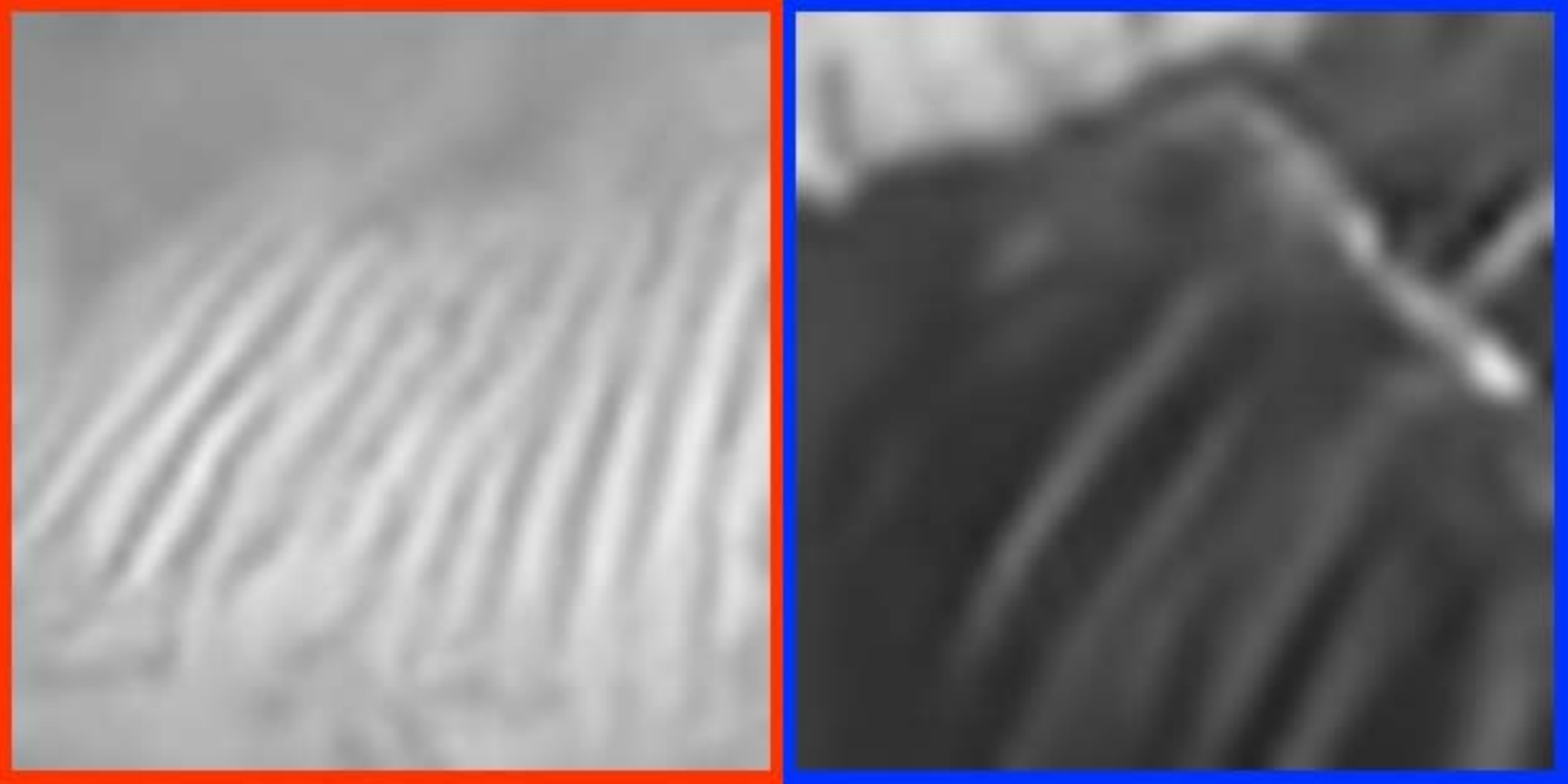} \\
				(a) clean  & (b) noisy   & (c) RAISR+C & (d) RAISR+N   & (e) PCN+N \\
				PSNR/SSIM  &  20.56/0.362   &  30.33/0.836 & 29.42/0.780   & 29.80/0.798  \\
			\end{tabular}
			\begin{tabular}{ccccccccc}
				\includegraphics[width=\swnine]{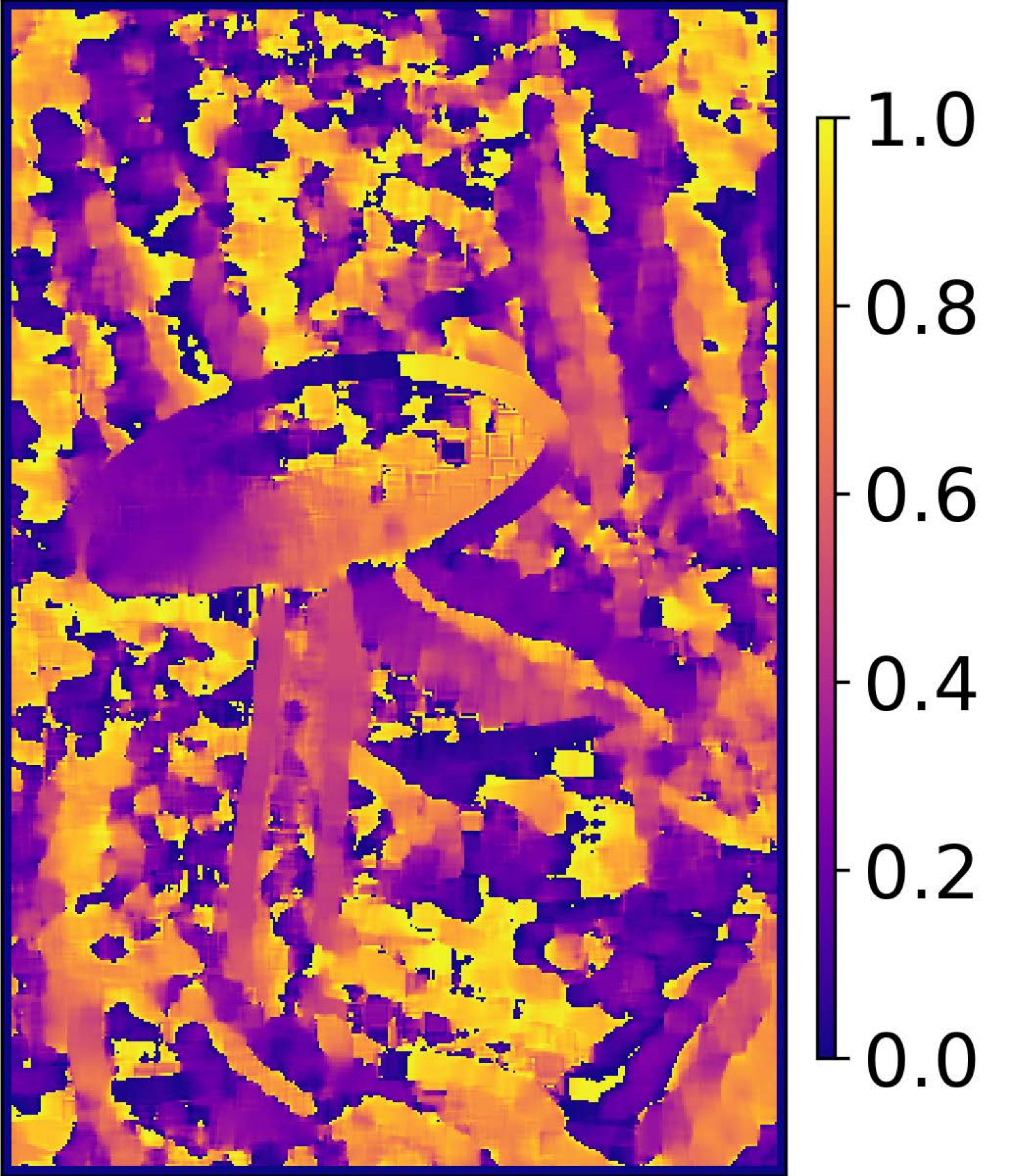} &
				\includegraphics[width=\swnine]{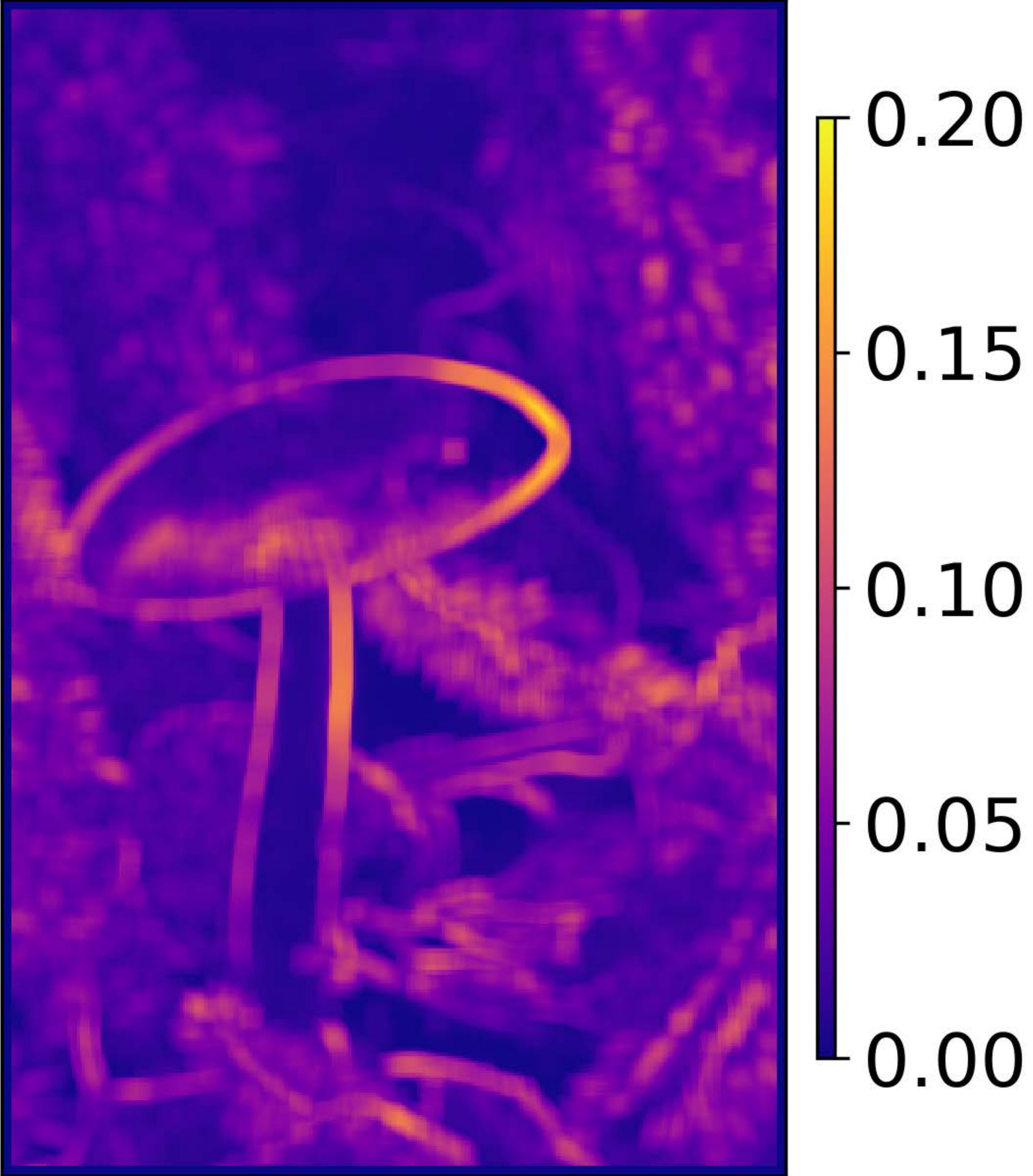} &
				\includegraphics[width=\swnine]{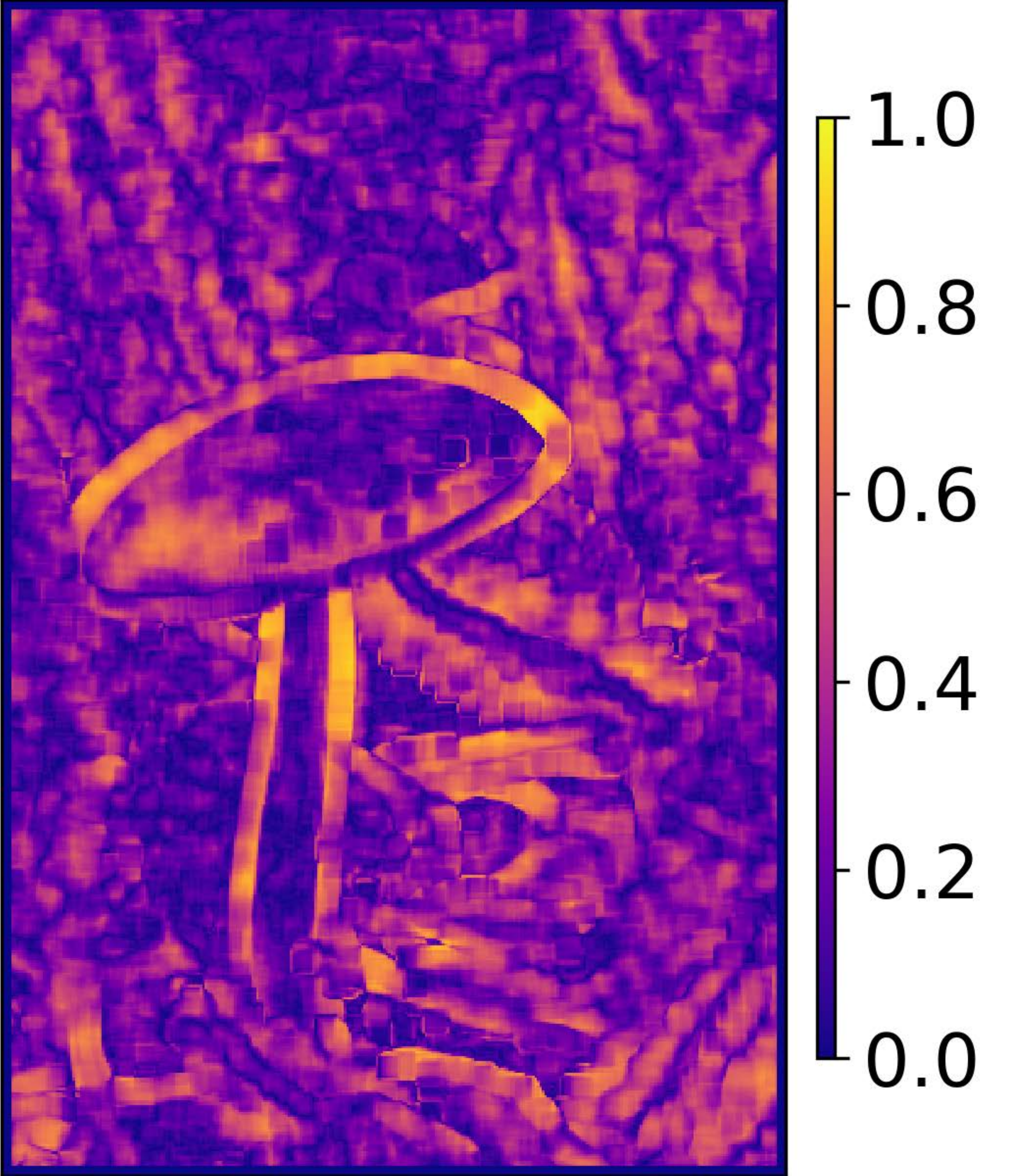} & %\\
				%\multicolumn{3}{c}{(f) RAISR+C }\\
				%\multicolumn{3}{c}{MSE }\\
				\includegraphics[width=\swnine]{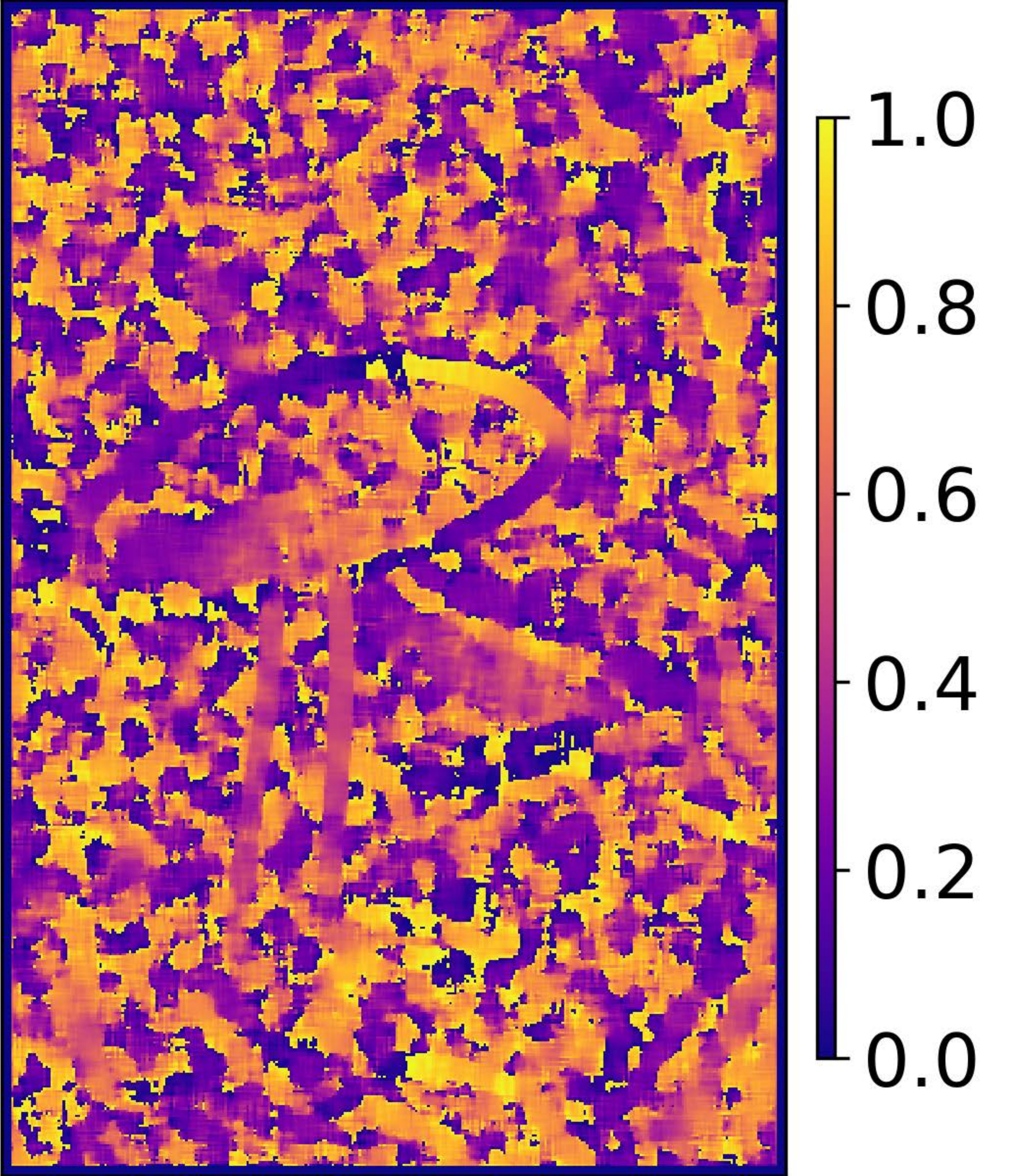} &
				\includegraphics[width=\swnine]{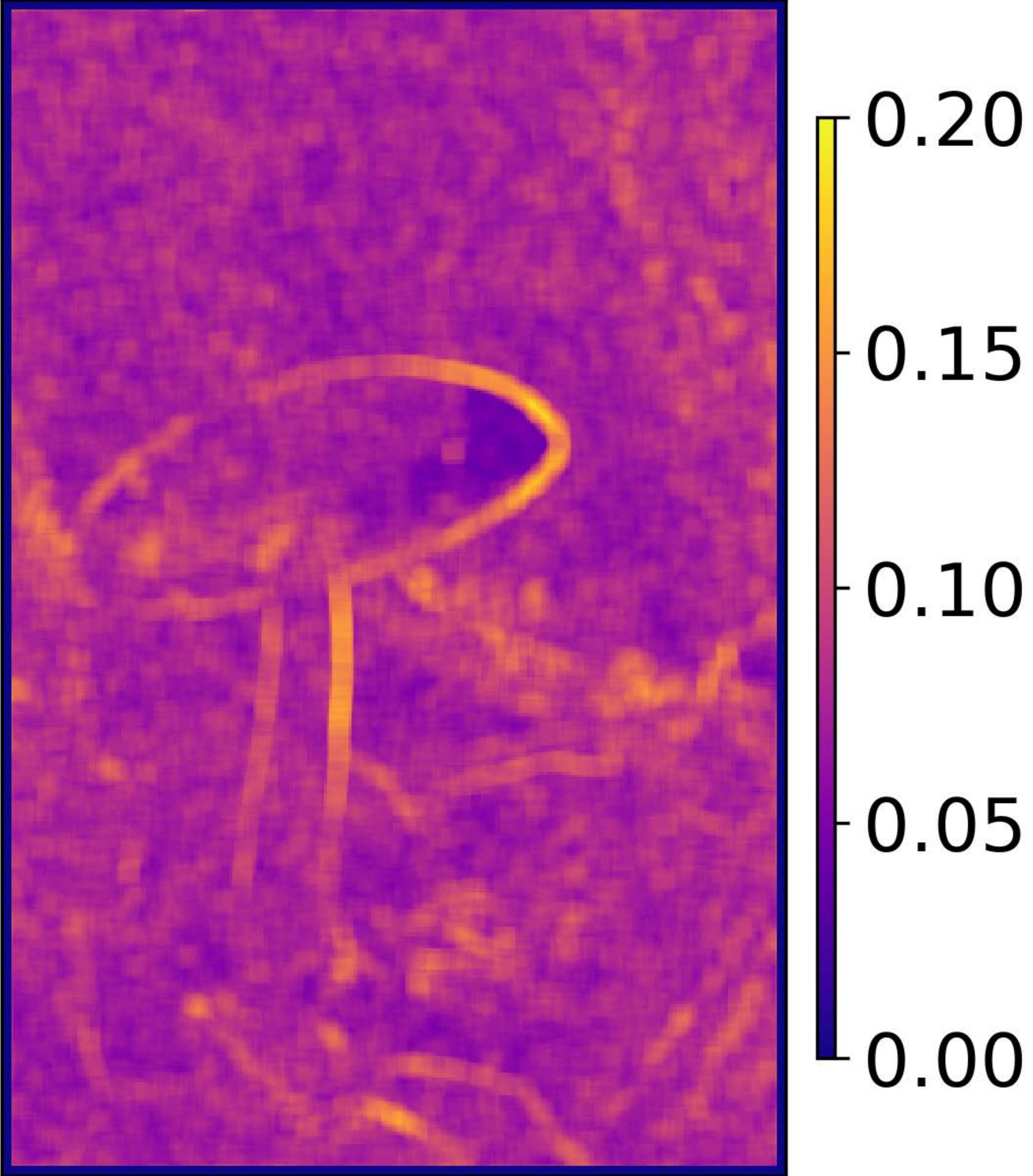} &
				\includegraphics[width=\swnine]{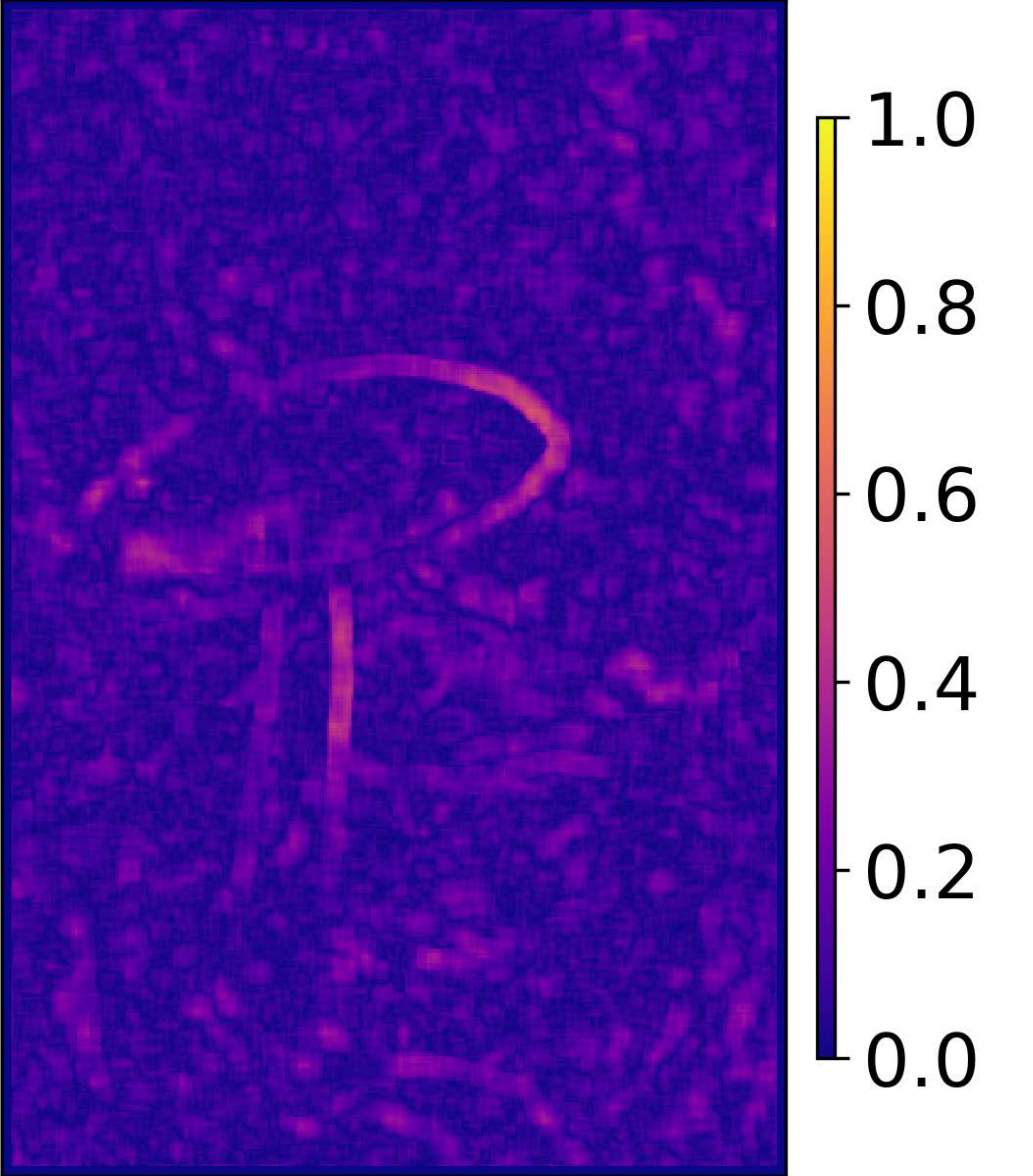} & %\\
				%\multicolumn{3}{c}{(g) RAISR+N }\\
				%\multicolumn{3}{c}{ 0.063 }\\
				\includegraphics[width=\swnine]{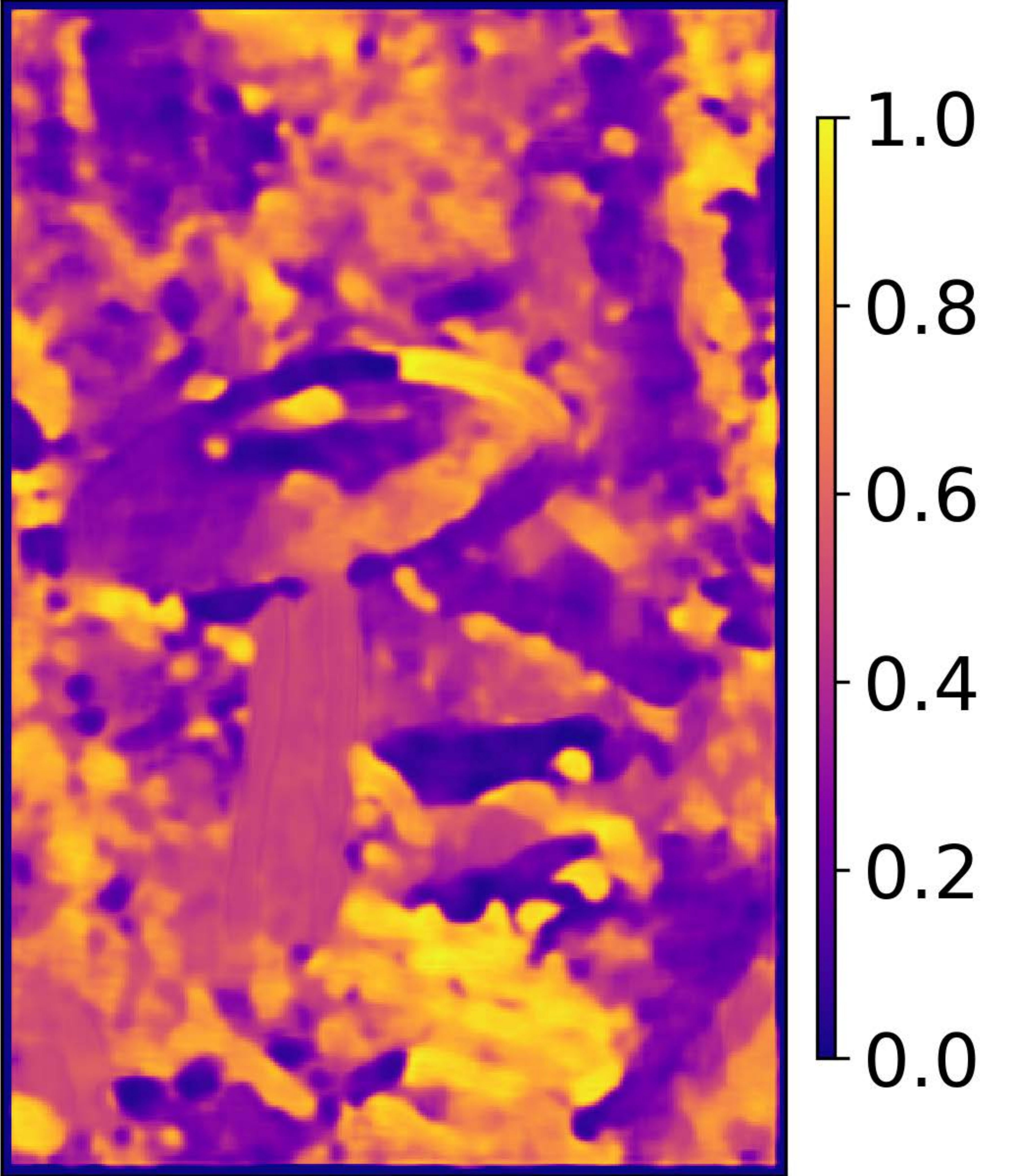} &
				\includegraphics[width=\swnine]{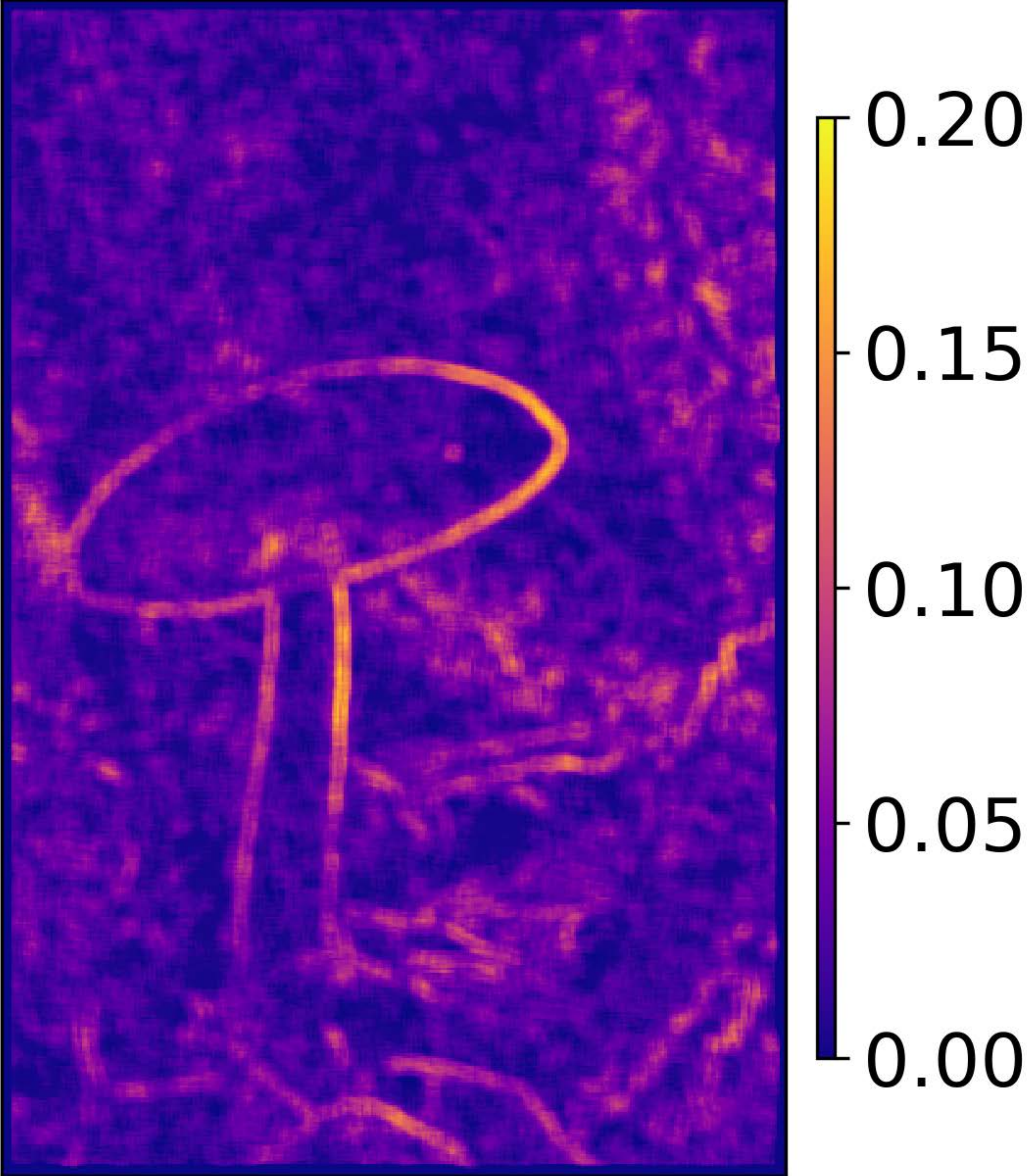} &
				\includegraphics[width=\swnine]{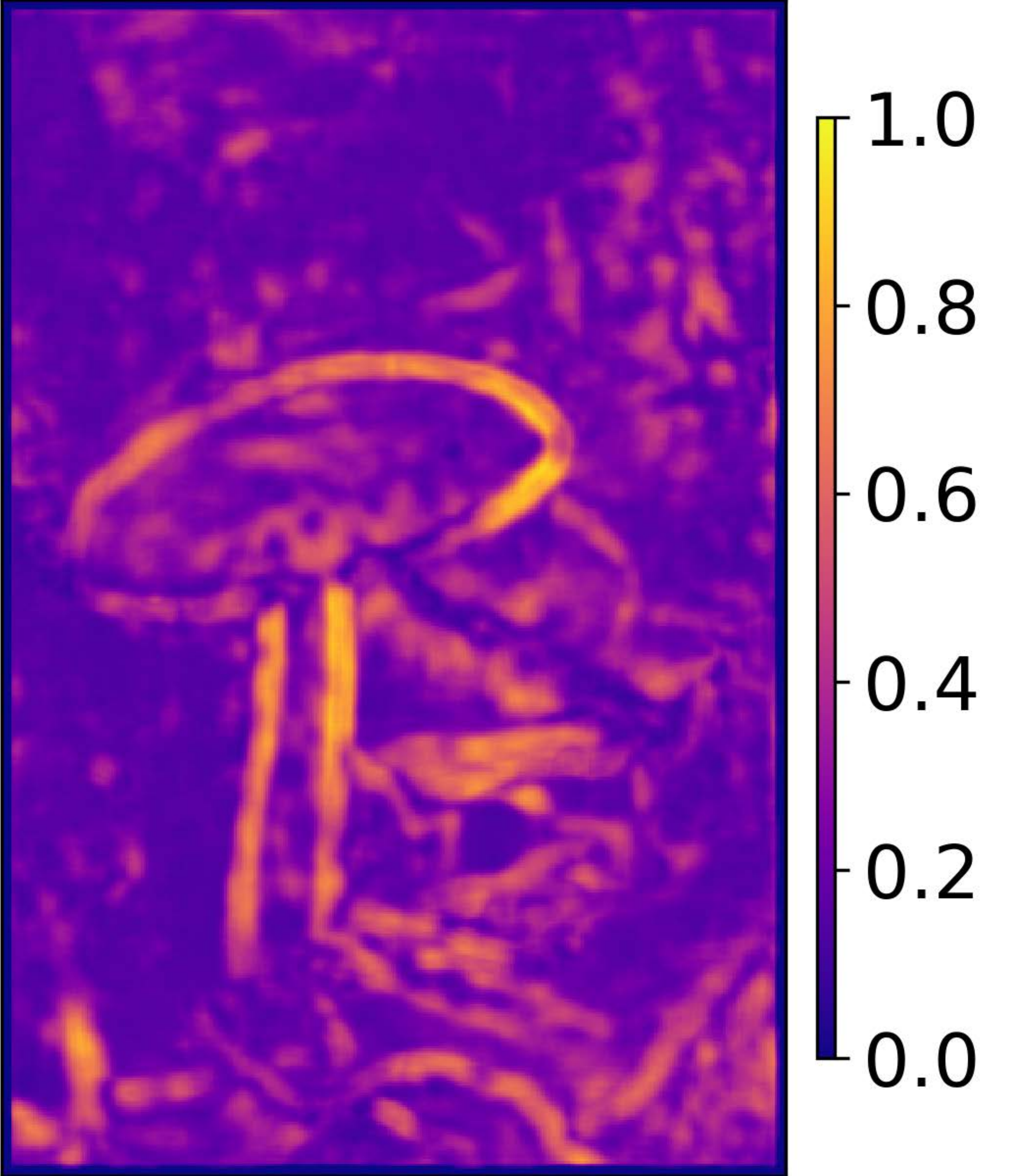} \\
				%\multicolumn{3}{c}{(h) PCN+N }\\
				%\multicolumn{3}{c}{ 0.029 }\\
				\multicolumn{3}{c}{(f) RAISR+C } & \multicolumn{3}{c}{(g) RAISR+N }  & \multicolumn{3}{c}{(h) PCN+N }\\
				\multicolumn{3}{c}{MSE } & \multicolumn{3}{c}{ 0.063 }  & \multicolumn{3}{c}{ 0.029 }\\
				
			\end{tabular}
		\end{center}
		%\vspace{-3mm}
		\caption{
			Top row: visual comparison by different classification methods on image ``0041" from BSD68 with noise $\sigma =25$. (a)-(e) are the cropped input, ground truth as well as outputs of different methods. As can be seen in the zoomed area, PCN+N and RAISR+C recover the sharpness and the direction of the stripe pattern. Bottom row: visualized gradient statistics comparison of the whole images produced by different classification methods. For each method, three figures from left to right denote the pixel-wise gradient statistics $\phi$, $\lambda$ and $\mu$. Directly using the RAISR gradient statistics under noises inaccurate pixel classifications. PCN produces a close estimate of the RAISR+C while RAISR+N is disrupted by the existence of noises.
		}
		\label{fig:effective}
	\end{figure*}
	\section{Ablation Study}
	In this section, several experiments are conducted with $\sigma = 25$ noises to validate the effectiveness of different components in the proposed network.
	%
	%	First, we compare the proposed network with RAISR~\cite{romano2016raisr} and KPN~\cite{mildenhall2018burst} which are the two most related previous works of our method.
	%	%
	%	Then, we validate that the proposed PCN can perform better than the eigenanalysis based on RAISR~\cite{romano2016raisr} when classifying pixels from noises.
	%	%
	%	We also compare the proposed networks with different estimated numbers of classes for $\phi$, $\lambda$ and $\mu$.
	%	%
	%	At last, we validate that the network will benefit more from the CSConv with fewer blocks or features in CSDN.

	%	\subsection{Comparison with RAISR~\cite{romano2016raisr} and KPN~\cite{mildenhall2018burst}}
	\subsection{Comparison with RAISR and KPN}
	As we discussed above, the most relevant works of the proposed network are RAISR and KPN.
	Whereas, both of them have limitations.
	Their spatially variant convolutions are directly applied to images and they are difficult to simultaneously remove severe noises and preserve textures which can be seen in \reffig{fig:raisrkpn} (e)(f).
	To make KPN deeper and have a fair comparison to CS-EDSR, we use the proposed PCN to predict the spatially variant kernels directly like KPN and replace CSConv by them in CS-EDSR which is denoted as KPN+.
	However, it performs worse than the proposed CS-EDSR in both \reffig{fig:raisrkpn}~(d) and \reftable{table:raisr}.
	The major reason is that many parameters need to be estimated from KPN, which makes it too difficult for an efficient network.
	For the proposed method, PCN only needs to classify the pixels into different classes which is much easier and it is more reasonable to utilize the divide-and-conquer strategy for image denoising by the proposed CSConv.
	According to \reftable{table:raisr}, the proposed CS-EDSR results in better average PSNR and SSIM.

	\begin{figure*}[!t]
		\begin{center}
			\begin{tabular}{cccccc}
				%\vspace{-4.5mm}
				\includegraphics[width=\swsix]{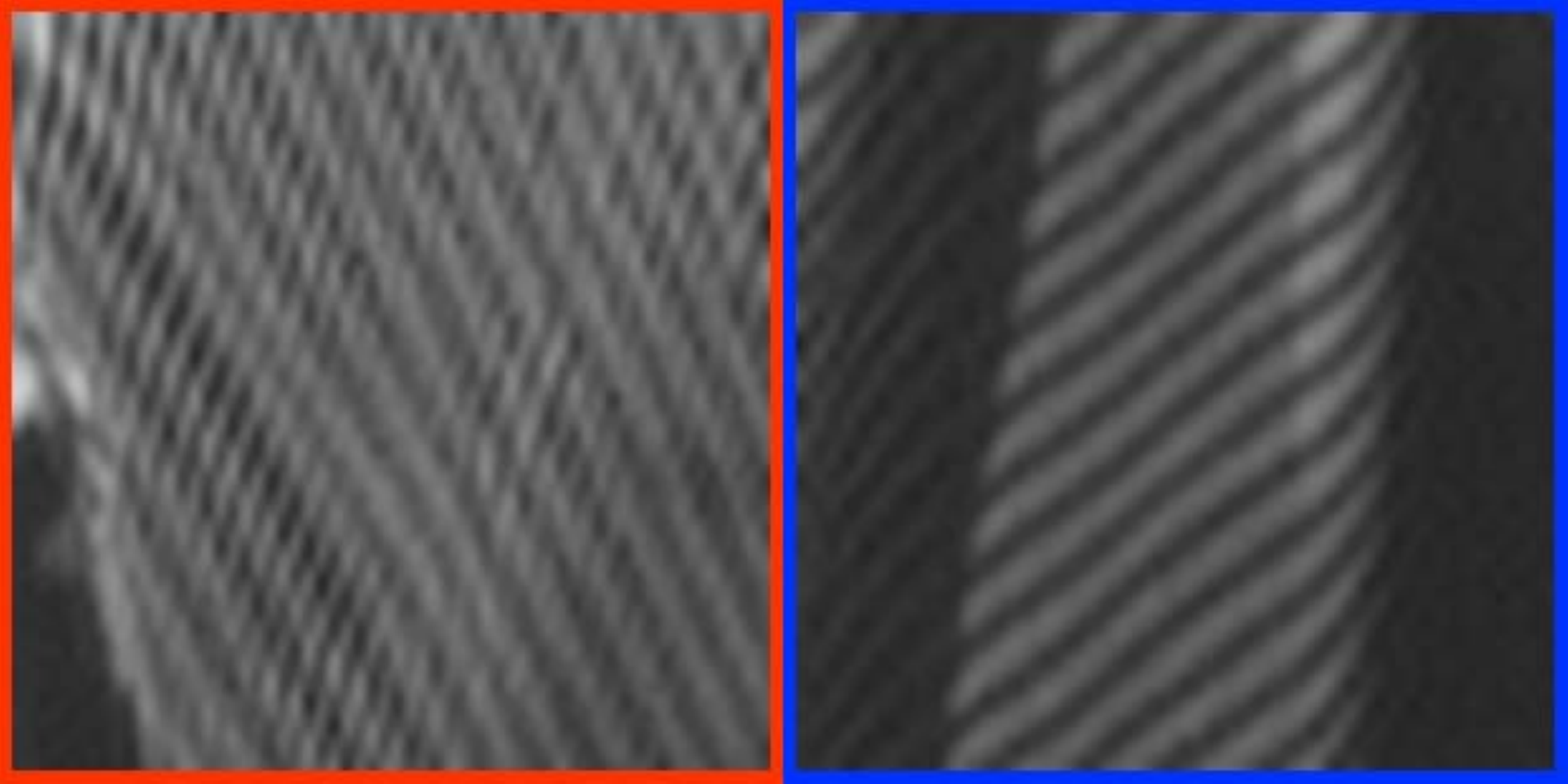} &
				\includegraphics[width=\swsix]{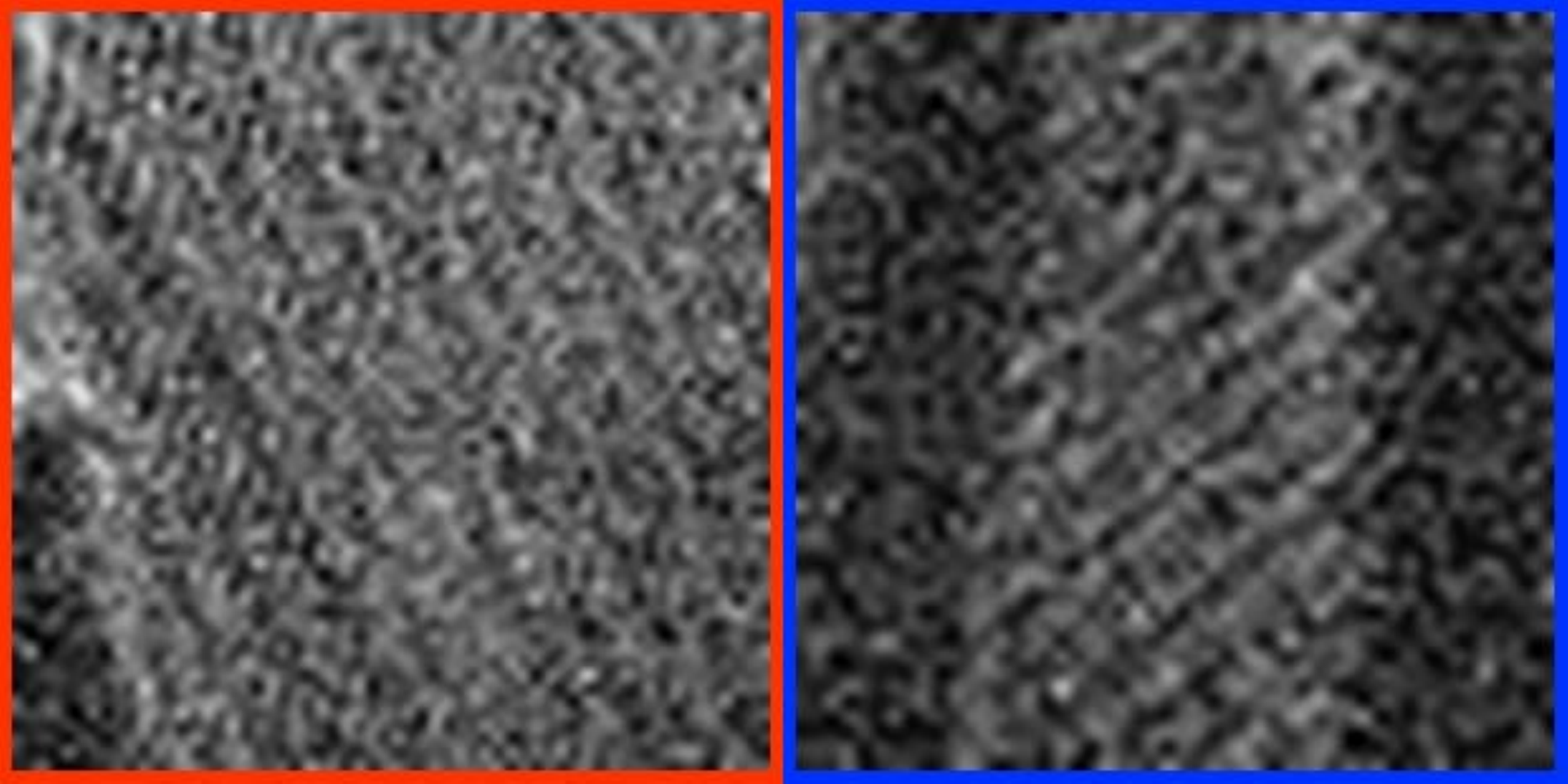} &
				\includegraphics[width=\swsix]{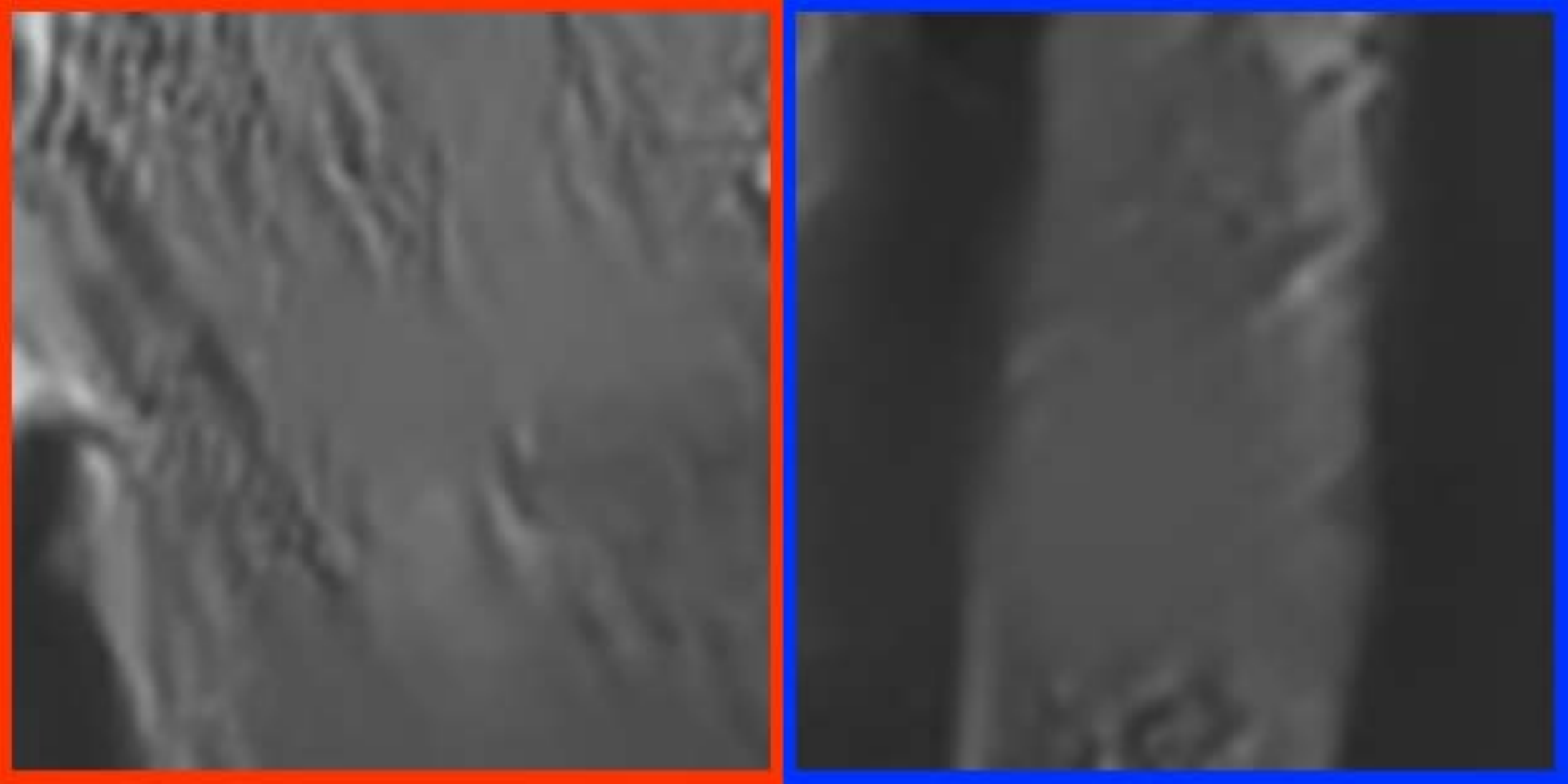} &
				\includegraphics[width=\swsix]{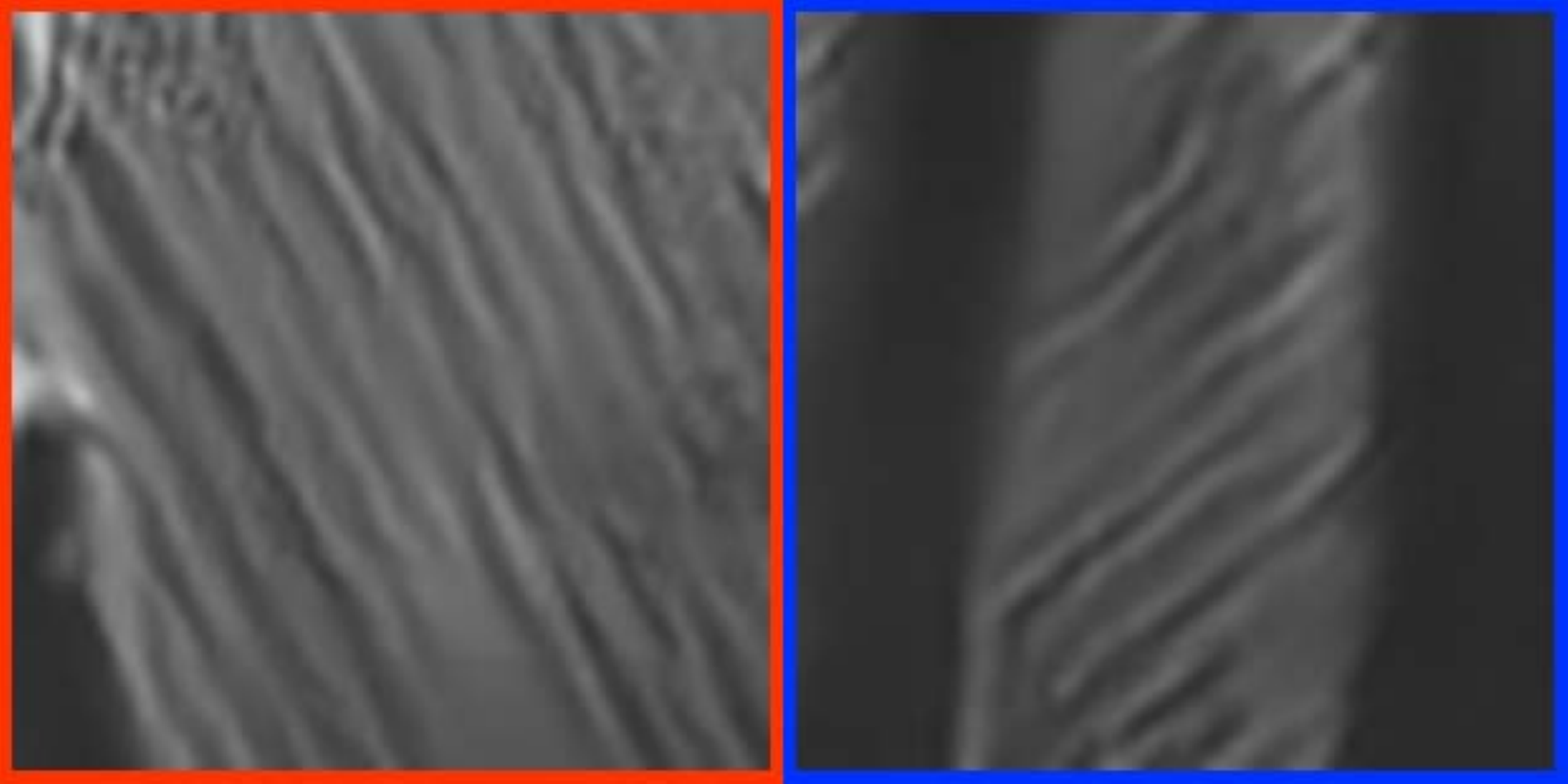} & \includegraphics[width=\swsix]{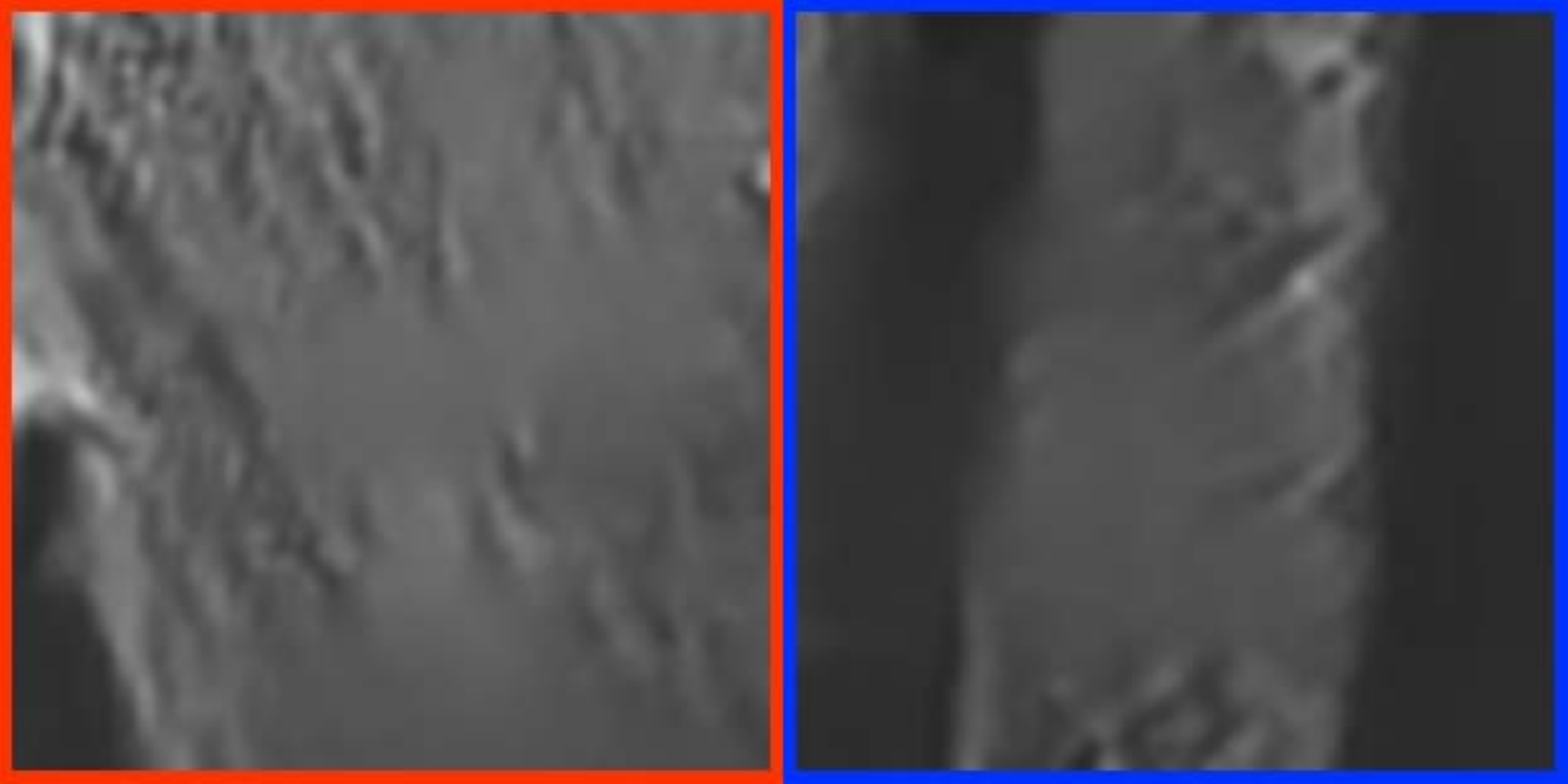} &
				\includegraphics[width=\swsix]{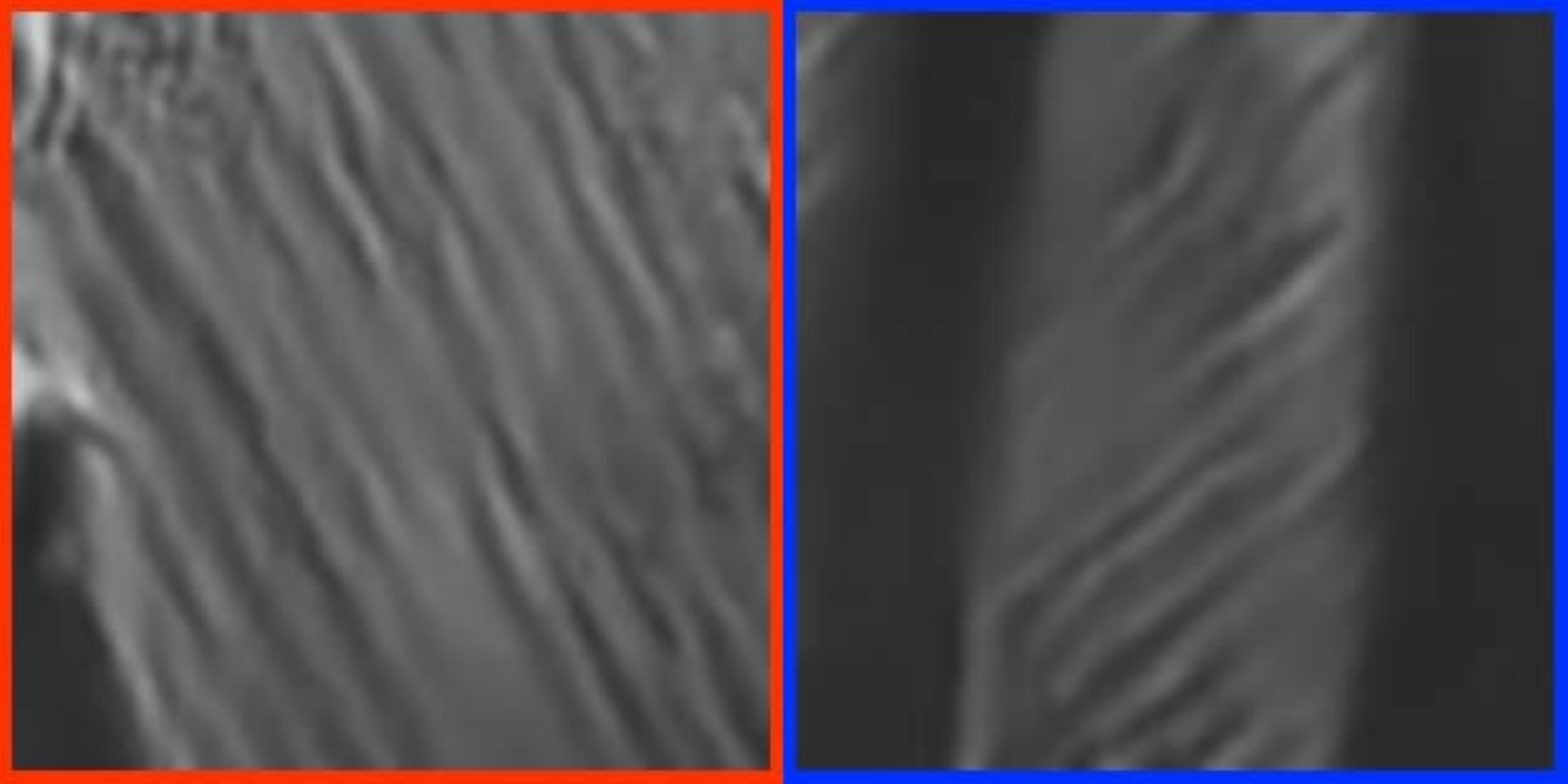}
				\\
				%\vspace{-1mm}
				(a) clean & (b) noisy & (c) $B_8F_{16}$ & (d) CS-$B_8F_{16}$ & (e) $B_{16}F_8$ & (f) CS-$B_{16}F_8$\\
				PSNR/SSIM & 20.42/0.395 & 29.28/0.839 & 29.82/0.867 & 28.83/0.839 & 29.58/0.861\\
				
				\includegraphics[width=\swsix]{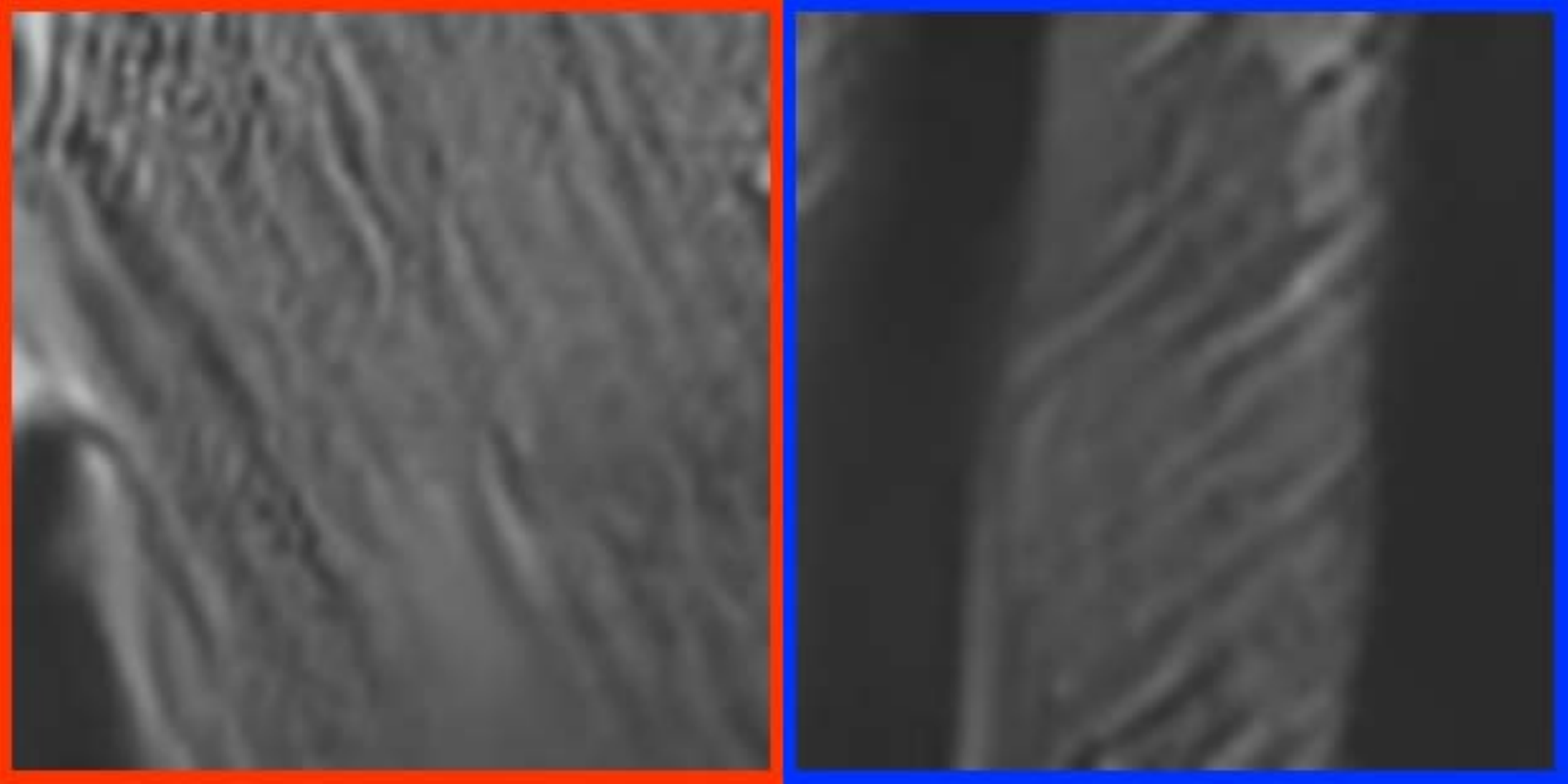} &
				\includegraphics[width=\swsix]{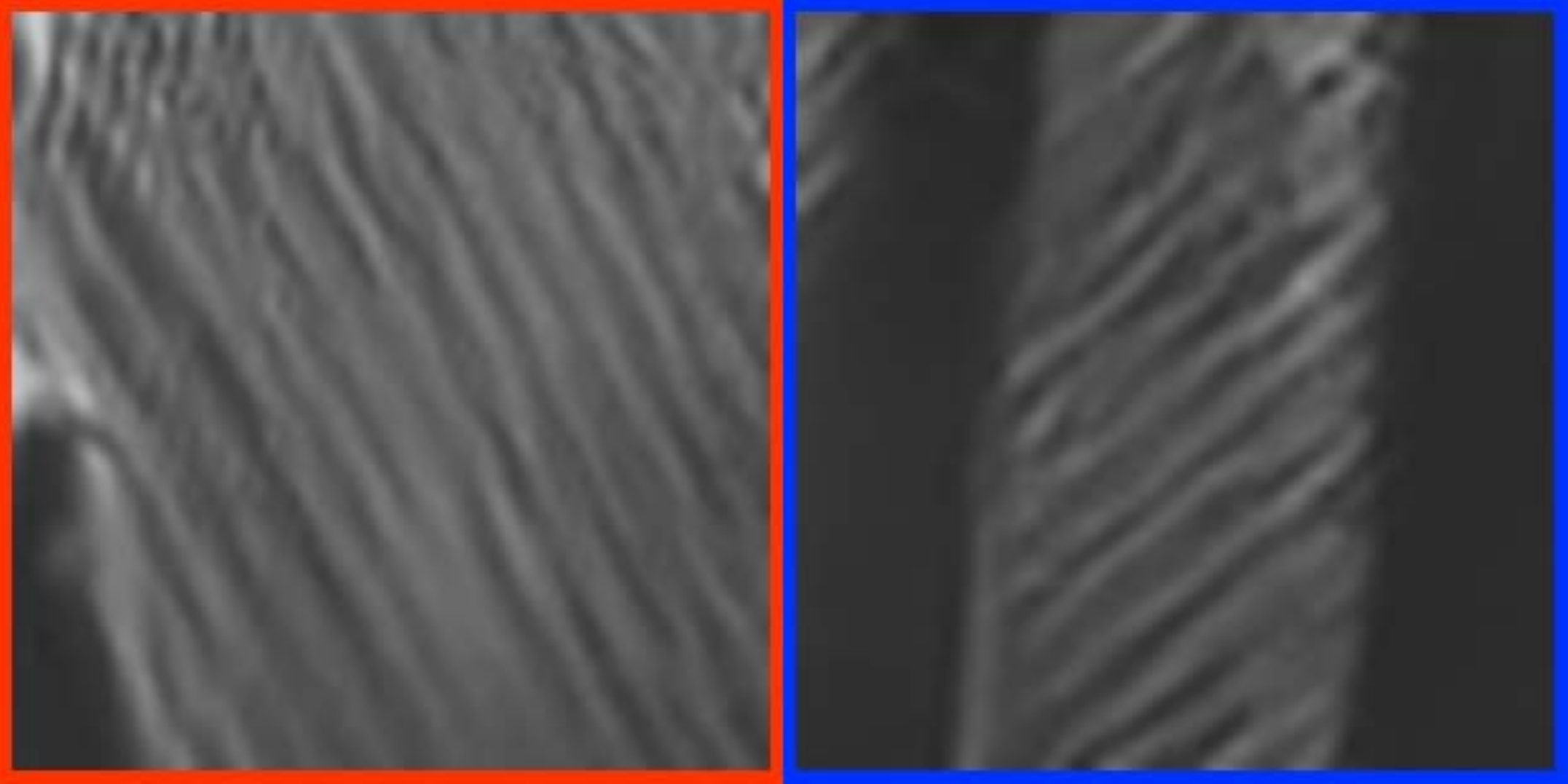} 
				&
				\includegraphics[width=\swsix]{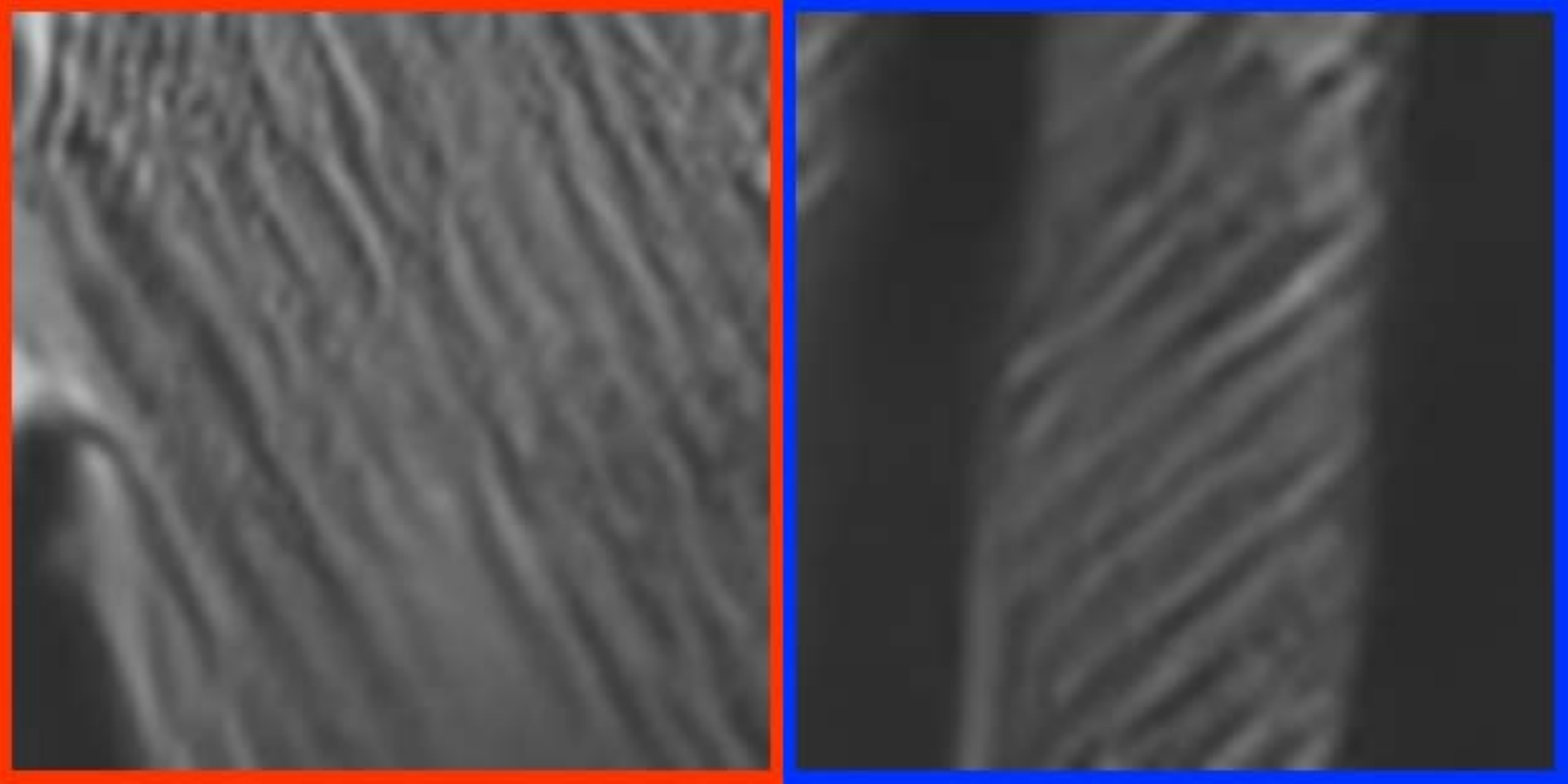} &
				\includegraphics[width=\swsix]{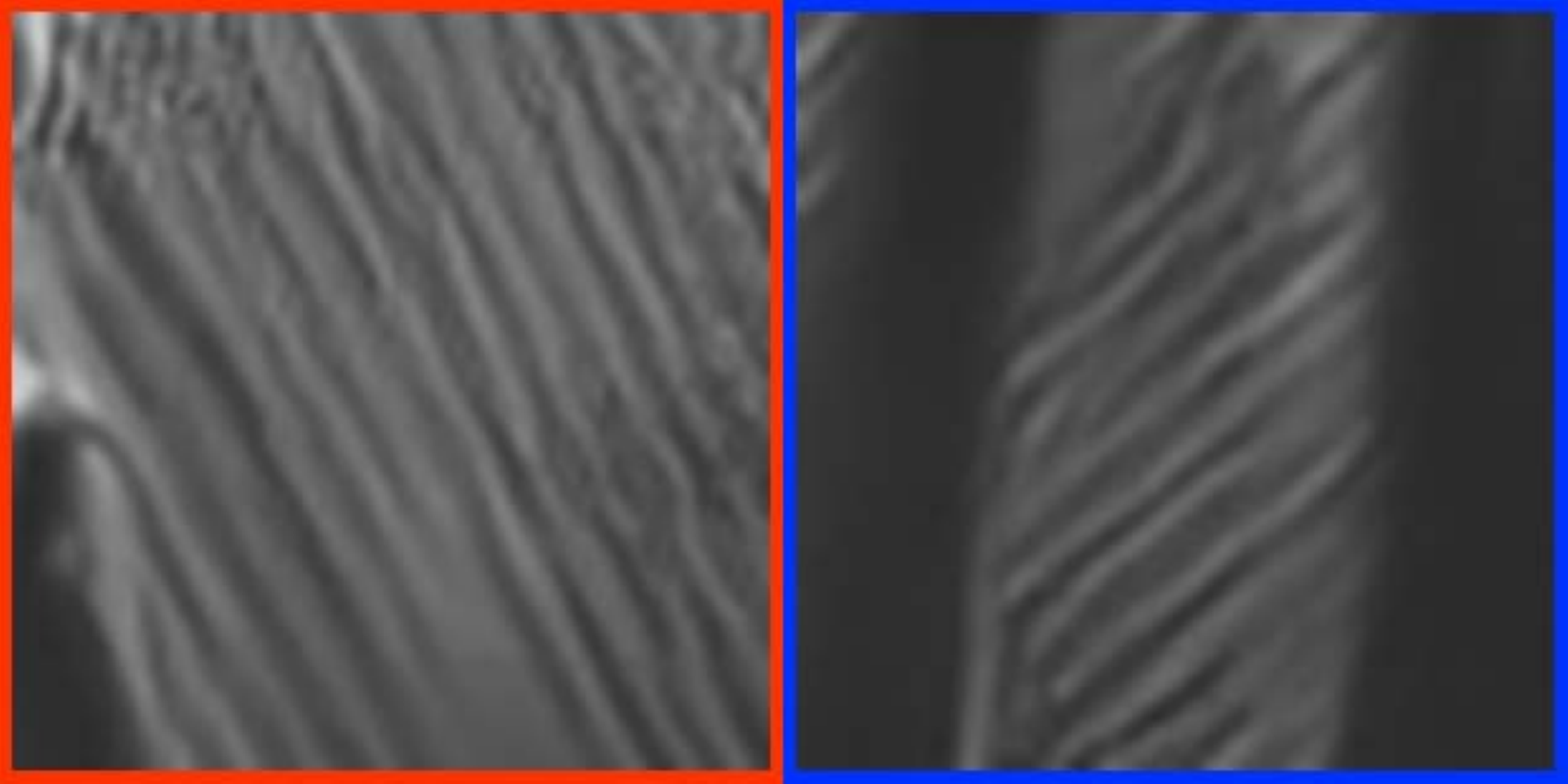} &
				\includegraphics[width=\swsix]{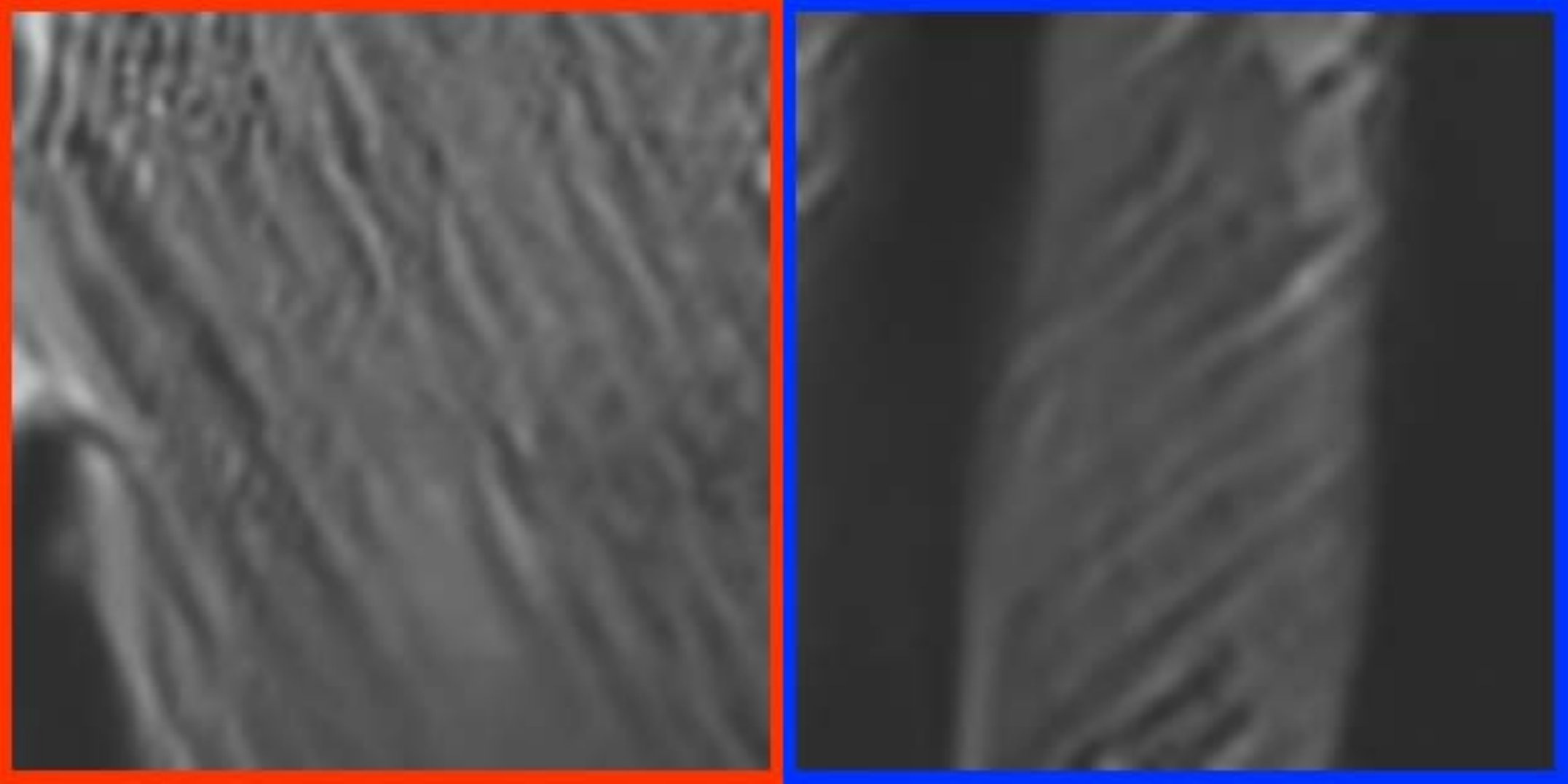} &
				\includegraphics[width=\swsix]{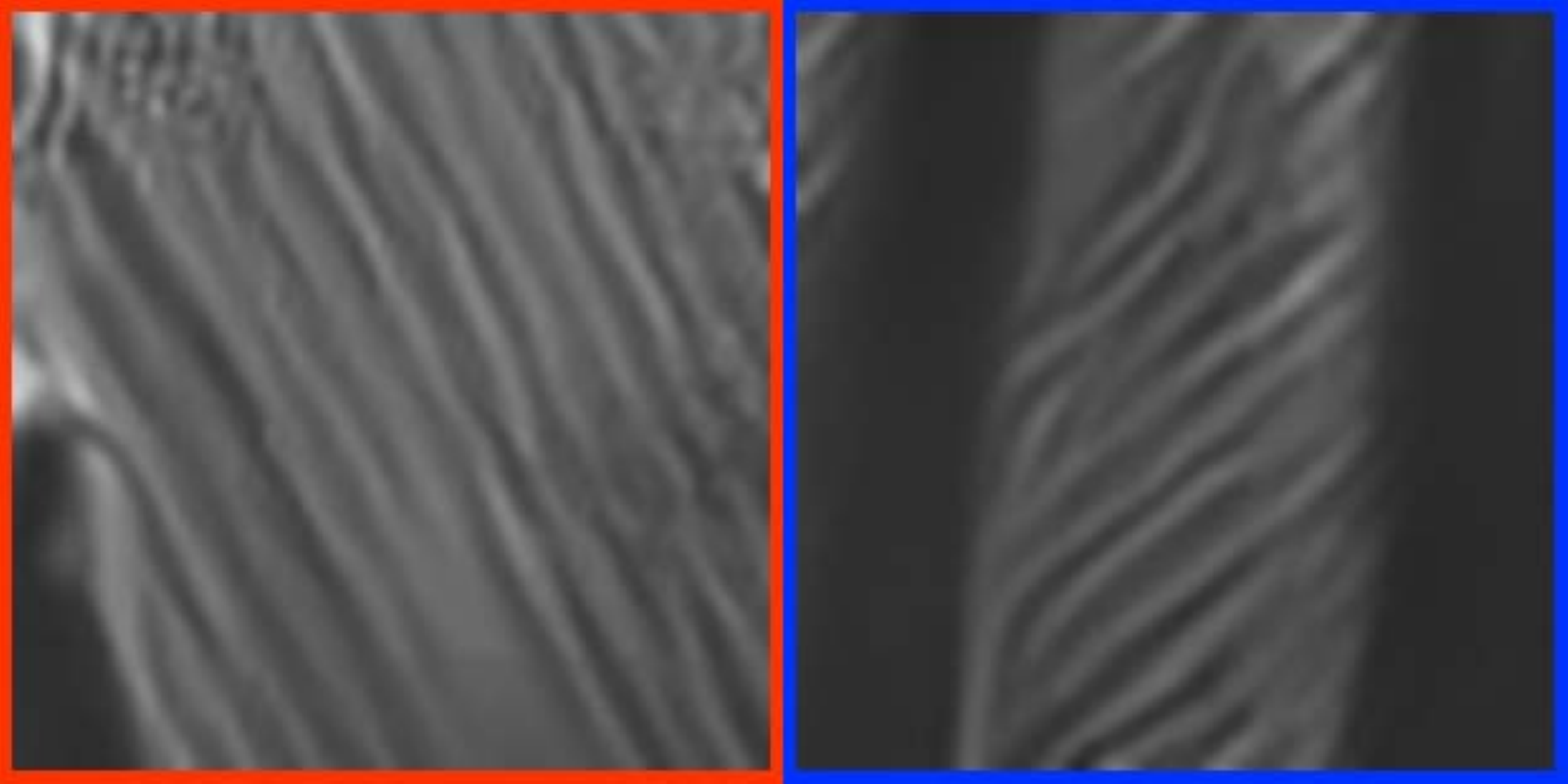}
				%\vspace{-1mm}
				\\
				(g) $B_{16}F_{16}$ & (h) CS-$B_{16}F_{16}$ & (i) $B_{16}F_{32}$ & (j) CS-$B_{16}F_{32}$ & (k) $B_{32}F_{16}$ & (l) CS-$B_{32}F_{16}$\\
				30.01/0.871 & 30.21/0.877& 30.23/0.878 & 30.29/0.879 & 30.16/0.873 & 30.28/0.879  \\
			\end{tabular}
			%\caption{Second subfigure} \label{subfig:upper-right}
		\end{center}
		%\vspace{-4mm}
		\caption{
			Visual comparison by CS-EDSR and EDSR with different numbers of blocks and features on image ``barbara" from Set12 with $\sigma =25$ noises.
			Model names with and without CS- denote as CS-EDSR and EDSR, respectively.
			$B_{i}F_{j}$ denotes the number of blocks and features as $i$ and $j$, respectively. 
			As can be seen in the zoomed areas, when using fewer blocks or features, CS-EDSR still produces relatively clear details, whereas the results of EDSR are smoother.
		}
		\label{fig:ablationblocks}
	\end{figure*}

\begin{table}[!t]
	\begin{center}
		
		\resizebox{\linewidth}{!}{
			\begin{tabular}{l|lll|lll} \hline \hline
				& \multicolumn{3}{c|}{EDSR}  & \multicolumn{3}{c}{CARN} \\
				Dataset  & PCN+N  & RAISR+C & RAISR+N & PCN+N & RAISR+C & RAISR+N  \\ \hline
				BSD68    & 29.76/0.826     & 30.22/0.847  & 29.42/0.814         & 29.65/0.820      & 29.90/0.838    & 29.36/0.812        \\
				Urban100 & 30.17/0.882     & 30.47/0.896   & 29.36/0.866        & 29.83/0.874     & 29.97/0.885   & 29.36/0.863         \\
				DIV2K    & 31.74/0.861     & 32.22/0.878   & 31.28/0.849       & 31.58/0.856     & 31.87/0.868   & 31.22/0.845       \\
				Set12    & 30.97/0.853     & 31.39/0.868   & 30.40/0.840        & 30.78/0.847     & 31.00/0.859   & 30.31/0.838        \\ \hline\hline
		\end{tabular}}
		\caption{
			Denoising performance (PSNR/SSIM) with $\sigma = 25$ noises by different pixel-wise classification.
			PCN+N denotes using PCN to classify pixels from noisy images.
			And RAISR+C and RAISR+N are using the eigenanalysis as to RAISR for pixel-wise classification from clean and noisy image, respectively.
			With clean image for classification, RAISR+C obviously performs the best.
			But PCN+N is better than RAISR+N.
		}
	\label{table:ablationclass}
	\end{center}
\end{table}

	\subsection{Effectiveness of PCN}
	The proposed PCN can estimate the image gradient statistics, which are used for pixel-wise classification, from the noisy image by utilizing the model capacity of the network.
	To validate the effectiveness of PCN, we also use exactly the same eigenanalysis as to RAISR to pixel-wisely regress $\phi$, $\lambda$ and $\mu$ from both the noisy and clean images which are denoted as RAISR+N and RAISR+C, respectively.
	And we treat the estimation from clean image as the ground truth.
	%
	%As can be seen in \reftable{table:ablationclass}, the mean squared error (MSE) is much lower by the proposed PCN.
	%
	
	We compare their estimation in terms of mean square error (MSE) in \reftable{table:mse}.
	PCN predictions are relatively robust to noises, whereas directly applying RAISR classification to the noisy image degrades significantly as the noise level increases.
	According to \reffig{fig:effective}, PCN can estimate gradient statistics more accurately than RAISR+N.
	
	We also conduct another experiment which uses the above three different methods to classify pixels and then uses the same CS-EDSR for denoising.
	According to \reffig{fig:effective} and \reftable{table:ablationclass}, RAISR+C can obviously achieve the best performance with given the clean image for pixel classification.
	But the proposed PCN can still recover more details and reach higher PSNR and SSIM than RAISR+N which demonstrates the effectiveness of PCN.

		\begin{table}[!t]
			\begin{center}
				
				%\vspace{-2mm}
				
				\resizebox{\linewidth}{!}{
					\begin{tabular}{l|lll|lll} \hline \hline
						& \multicolumn{3}{c|}{EDSR}  & \multicolumn{3}{c}{CARN} \\
						Dataset & $\phi_4\lambda_2\mu_2$  & $\phi_8\lambda_3\mu_3$  & $\phi_{16}\lambda_6\mu_6$ & $\phi_4\lambda_2\mu_2$  & $\phi_8\lambda_3\mu_3$  & $\phi_{16}\lambda_6\mu_6$  \\ \hline
						BSD68  & 29.67/0.823  & 29.76/0.826         & 28.74/0.801    
						& 29.56/0.817 & 29.65/0.820   & 29.48/0.818    \\
						Urban100  & 29.86/0.879 & 30.17/0.882       & 29.26/0.860 
						& 29.49/0.870 & 29.83/0.874   & 29.27/0.869    \\
						DIV2K    & 31.59/0.858   & 31.74/0.861         & 30.76/0.845 
						& 31.43/0.852 & 31.58/0.856   & 31.38/0.853    \\
						Set12  & 30.66/0.850   & 30.97/0.853        & 29.67/0.837 
						& 30.58/0.844  & 30.78/0.847   & 30.33/0.843  \\ \hline\hline
				\end{tabular}}
			\caption{
				Denoising performance (PSNR/SSIM) with $\sigma = 25$ noises with different number of classes.
				$\phi_i\lambda_j\mu_k$ denotes $\phi$, $\lambda$ and $\mu$ are divided into $i$, $j$ and $k$ classes, respectively.
				The default setting in our network is $\phi_8\lambda_3\mu_3$.
				%
				%					and the improvement is marginal with more classes.
			}
			\label{table:classnumber}	
			\end{center}
		\end{table}
		%\setlength{\tabcolsep}{1.4pt}
		%\vspace{-4mm}

	\subsection{Different Number of Classes}
	In this subsection, some experiments are conducted by comparing the denoising performance of both CS-CARN and CS-EDSR with different number of pixel classes.
	$\phi$, $\lambda$ and $\mu$ are divided into $M_{\phi}=8$, $M_{\lambda}=3$ and $M_{\mu}=3$ classes in all the aforementioned experiments.

	In \reftable{table:classnumber}, we also consider $M_{\phi}=16$, $M_{\lambda}=6$, $M_{\mu}=6$ as well as $M_{\phi}=4$, $M_{\lambda}=2$, $M_{\mu}=2$.
	Too many or too few classes result in inferior performance for both CS-CARN and CS-EDSR, possibly because too few classes limit the expressive power of the CSConv, and too many classes increase the difficulty for PCN to accurately classify pixels. Consequently, we select to use $M_{\phi}=8$, $M_{\lambda}=3$ and $M_{\mu}=3$ in the proposed network.

	\subsection{Different Number of Blocks and Features}
\begin{figure}[!t]\footnotesize
		\begin{center}
			\begin{tabular}{c}
				\includegraphics[width=\linewidth]{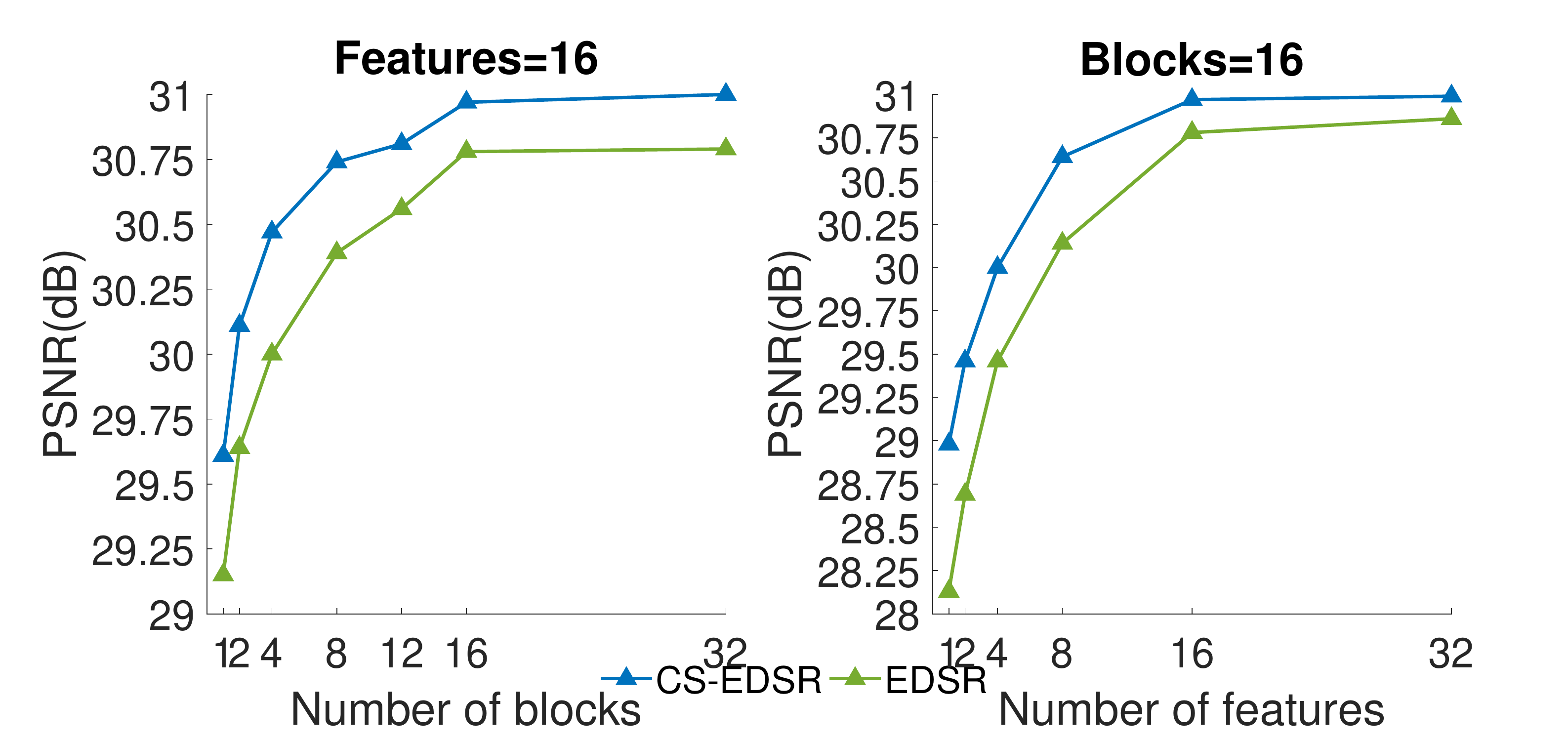} \\
			\end{tabular}
		\end{center}
		%\vspace{-5mm}
		\caption{
			Denoising performance (PSNR) of EDSR and CS-EDSR with different numbers of blocks and features on Set12 with $\sigma =25$ noises.
			Blue and green lines denote CS-EDSR and EDSR models respectively.
			With a smaller model, the performance of EDSR drops faster than the proposed CS-EDSR.
		}
		\label{fig:ablationplot}
	\end{figure}

	To further validate the helpfulness of the proposed CSConv for network efficiency, we compare EDSR and CS-EDSR with different numbers of blocks and features.
	As can be seen in \reffig{fig:ablationplot}, the performance of EDSR drops faster than that of CS-EDSR with fewer residual blocks or features.
	And EDSR cannot effectively remove noises when the network is small according to \reffig{fig:ablationblocks}.
	CS-EDSR with reduced channels or depths yields uncompromised visual results compared with EDSR64 while reducing those of EDSR causes significant performance drop in \reffig{fig:ablationblocks}.
	And this demonstrates that CSConv has more advantages when the model size is smaller which is more appropriate for mobile devices. 
	 The advantage of CSConv is marginal with more features, possibly because the network already has enough capacity.

	\section{Conclusion}
	In this paper, we utilize the gradient statistics to take a divide-and-conquer scheme on image denoising.
	A deep neural network pipeline is proposed to first classify pixels into classes, then perform image denoising with a small network using the proposed Class Specific Convolution (CSConv).
	CSConv, which applies dedicated weights for different pixel classes, can replace convolution layers in state-of-the-art denoising networks.
	With a smaller number of features, the proposed network with CSConv can reduce the computational cost while maintaining the denoising performance.
	The proposed method is evaluated on additive Gaussian denoising benchmarks and obtains competitive denoising performance relative to state-of-the-art methods with less computation.

	\bibliography{egbib}
\end{document}